\documentclass{amsart}
\usepackage[top=1.2in, bottom=1.2in, left=1.0in, right=1.0in]{geometry}

\usepackage[utf8]{inputenc}
\usepackage{amsmath}
\usepackage{amsfonts}
\usepackage{amssymb}
\usepackage[makeroom]{cancel}
\usepackage[titletoc,toc]{appendix}
\usepackage{tcolorbox}
\usepackage{color}
\usepackage{hyperref}
\usepackage{cleveref}
\usepackage{tikz}
\usepackage[ruled]{algorithm2e}
\usepackage[colorinlistoftodos]{todonotes}
\usepackage{multirow}
\usepackage{natbib}

\definecolor{darkblue}{rgb}{0.0,0.0,0.5}
\hypersetup{colorlinks,breaklinks,
            linkcolor=darkblue,urlcolor=darkblue,
            anchorcolor=darkblue,citecolor=darkblue}

\newcommand{\lr}[1]{\left(#1\right)} 
\newcommand{\ve}[1]{\text{\boldmath$#1$}} 
\newcommand*\dint{\,\mathrm{d}}
\newcommand{\partDeriv}[2]{\frac{\partial #1}{\partial #2}}

\renewcommand{\lr}[1]{\left( #1 \right)}
\newcommand{\abs}[1]{\left|#1\right|}
\newcommand{\norm}[1]{\left\lVert#1\right\rVert}

\definecolor{mygray}{RGB}{82,87,110}

\newcommand{\x}{\ve{x}}
\newcommand{\sx}{x}
\newcommand{\y}{\ve{y}}

\newcommand{\mivel}{\ve{u}}
\newcommand{\mipres}{p}
\newcommand{\miconc}{c}
\newcommand{\miconcdim}{\tilde{c}}

\newcommand{\miradius}{r}
\newcommand{\miradiusdim}{\tilde{r}}
\newcommand{\partvol}{v}
\newcommand{\partd}{q}

\newcommand{\fibset}{\mathrm{M}}
\newcommand{\mifibset}{\mathrm{M}_{\midmn}}
\newcommand{\fib}{m}
\newcommand{\allfib}{\fib\in\fibset}
\newcommand{\allmifib}{\fib\in\mifibset}

\newcommand{\fibcentdim}{\tilde{\x}^{\fib}}
\newcommand{\miradiusfib}{\miradius^{\fib}}
\newcommand{\miradiusfibdim}{\miradiusdim^{\fib}}

\newcommand{\velv}{\overline{\ve{U}}}
\newcommand{\svelv}{\overline{U}}
\newcommand{\conc}{C}
\newcommand{\concflux}{J_{\conc}}

\newcommand{\pres}{P}

\newcommand{\pack}{\rho}
\newcommand{\conctot}{\mathcal{C}}
\newcommand{\ads}{k}
\newcommand{\adsdim}{\tilde{\ads}}
\newcommand{\diff}{D}
\newcommand{\visc}{\mu}
\newcommand{\sPerm}{\mathcal{K}}
\newcommand{\perm}{\boldsymbol{\mathcal{K}}}
\newcommand{\sDeff}{\mathcal{D}}
\newcommand{\deff}{\boldsymbol{\mathcal{D}}}
\newcommand{\normal}{\ve{n}}

\newcommand{\area}{\mathcal{A}}
\newcommand{\efficiency}{E}
\newcommand{\dhc}{H}
\newcommand{\dhcsens}{S_{\dhc}}
\newcommand{\pressens}{S_{\Delta\pres}}
\newcommand{\velsens}{S_{\velv}}
\newcommand{\timesens}{S_{T}}
\newcommand{\dmn}{\Omega}
\newcommand{\dmnf}{\dmn_f}
\newcommand{\dmns}{\dmn_s}
\newcommand{\bdrys}{\partial\dmns}

\newcommand{\dmndim}{\tilde{\dmn}}
\newcommand{\dmnfdim}{\tilde{\dmn}_f}
\newcommand{\dmnsdim}{\tilde{\dmn}_s}
\newcommand{\bdrysdim}{\partial\tilde{\dmn}_s}
\newcommand{\bdrysfib}{\bdrys^{\fib}}
\newcommand{\bdrysfibdim}{\bdrysdim^{\fib}}
\newcommand{\midmn}{w}
\newcommand{\midmnf}{\midmn_f}
\newcommand{\midmns}{\midmn_s}
\newcommand{\mibdry}{\partial \midmn}
\newcommand{\mibdrys}{\partial \midmns}
\newcommand{\mibdryf}{\partial \midmnf}
\newcommand{\mibdrysfib}{\mibdrys^{\fib}}
\newcommand{\miBdryF}{\mibdryf(\x,t)}
\newcommand{\miBdryS}{\mibdrys(\x,t)}
\newcommand{\miBdrySFib}{\mibdrysfib(\x,t)}

\newcommand{\miDmn}{\midmn(\x,t)}
\newcommand{\miDmnF}{\midmnf(\x,t)}
\newcommand{\miDmnS}{\midmns(\x,t)}

\newcommand{\coefA}{\alpha}
\newcommand{\coefB}{\beta}
\newcommand{\coefC}{\gamma}
\newcommand{\coefD}{\zeta}
\newcommand{\coefE}{\eta}

\crefname{figure}{Figure}{Figures}
\Crefname{figure}{Figure}{Figures}
\crefname{section}{Section}{Sections}
\crefname{appendix}{Appendix}{Appendices}
\crefname{table}{Table}{Tables}

\usepackage[normalem]{ulem}

\begin{document}
\title{The influence of porous media microstructure on filtration}

\author{G.~Printsypar}
\author{M.~Bruna}
\author{I.~M. Griffiths}

\address{Mathematical Institute, University of Oxford, Oxford, OX2 6GG, UK}
\begin{abstract} 
We investigate how a filter media microstructure influences filtration performance. We derive a theory that generalizes classical multiscale models for regular structures to account for filter media with more realistic microstructures, comprising random microstructures with polydisperse unidirectional fibres. Our multiscale model accounts for the fluid flow and contaminant transport at the microscale (over which the media structure is fully resolved) and allows us to obtain macroscopic properties such as the effective permeability, diffusivity, and fibre surface area. As the fibres grow due to contaminant adsorption this leads to contact of neighbouring fibres. We propose an agglomeration algorithm that describes the resulting behaviour of the fibres upon contact, allowing us to explore the subsequent time evolution of the filter media in a simple and robust way. We perform a comprehensive investigation of the influence of the filter-media microstructure on filter performance in a spectrum of possible filtration scenarios.
\end{abstract}
\maketitle

\section{Introduction}

Filtration of contaminant out of a fluid is vital for many industrial applications. Filtration technology is used in air conditioning and purifying systems, cars, vacuum cleaners, water treatment and food industries to name a few. Filtration in these applications operates under the same principles \citep{Filtration}. Contaminated fluid, such as air or water, is transported through a porous material, the filter media. As contaminants pass through the filter media, they come into contact with the surface of the porous media and adhere, and as a result, a cleaner fluid is produced. Filtration processes can be classified using four main characteristics: the transport mechanism, the operational set-up, the adsorption mechanisms and the filter media type. 

\subsubsection*{Transport mechanisms and operational set-ups}
The transport of contaminants though filter media can be facilitated via advection, diffusion and osmosis. In this work, we focus on the first two mechanisms. Depending on the transport mechanisms and the objectives of the filtration, the process can have different operational set-ups. A \emph{dead-end} set-up, when the fluid flow is perpendicular to the filter media, is used in advection-dominated filtration, while a \emph{cross-flow} set-up, when the fluid flow is parallel to the filter media, is commonly used for diffusion-dominated filtration. Moreover, when advection is present, the filtration can occur under a constant flow rate or a constant pressure drop. Some examples of applications that employ a constant flow rate are the air filters used in vacuum cleaners and air-conditioning systems \citep{Fisk2002}. Filtration regimes in which a constant pressure drop is applied occur in pharmaceutical and biotechnology industries, see, for example, \cite{CHEN201532, GOLDRICK2017138}. One of the aims of this work is to discuss the mathematical modelling and to investigate numerically different transport mechanisms and set-ups.

\subsubsection*{Adsorption mechanisms}
The adsorption mechanisms that usually act during the filtration process are diffusion, interception, impaction and gravitational settling. In addition, adsorption can be enhanced by, for example, electrostatic forces and chemical treatment of the filter media. In this paper, we ignore the enhanced adsorption mechanisms and account for the standard mechanisms through a single so-called adsorption coefficient. For more discussions on the adhesive forces acting on the contaminant particles and their quantification, see, for example, \cite{Brown1993} and \cite{Baron2001}.

\subsubsection*{Filter media types} 
Contaminant adsorption occurs at the pore level, or microscale, of the filter media. Therefore, a natural question that arises is how the filtration performance is affected by the microstructure of the filter media, that is, by the filter media type. The second aim of this paper is to investigate the influence of the microstructure on the filtration performance. 

The effect of the filter media type has been investigated in different studies using a \textit{microscale approach} with a fully resolved microstructure of filter media \citep[see, for example,][and references therein]{Fotovati2010, SAMBAER2012299, filterdict2013, Robinson, Li2016, porechem}. Some of these studies track each particle individually using a Lagrangian approach, while others treat the contaminant as a continuum, which is possible if the particles are sufficiently small in comparison with the fibre size. \citet{filterdict2013} evolve the microstructure as time progresses, while the other studies mentioned above consider only the initial filtration, namely before the adsorbed contaminants begin to influence the porous-media microstructure. In general, the microscale approach provides detailed information about the filtration process, but it is computationally very expensive. Using this approach, we can only consider a small representative volume of the filter, and so this does not provide us with the information about the behaviour of the entire filter media.  Even if we can resolve the whole thickness, due to the computational cost only a very limited number of simulations can be performed, which makes the microscale approach unsuitable for comprehensive studies with different kinds of microstructures. 

Filtration problems are also commonly modelled using a \textit{macroscale approach} \citep{Lakdawala2010,Manikantan2013,krupp2017scaling}. Here, the filter is modelled as a continuum and its characteristics are accounted for via empirical macroscopic parameters. Macroscale models are popular because one can relatively cheaply simulate the whole filtration process and various operational set-ups. Hence, unlike microscopic models, they are suitable for predictive studies. On the other hand, studying different types of filter media using this approach would require supplementing the simulations with experimental measurements, which are time-demanding and expensive to carry out. Hence, the macroscale approach is impractical for such a study.

A \textit{multiscale approach} combines the advantages of both micro- and macroscale methods. Starting from a microscale model, the multiscale approach uses an upscaling method to derive a simple model at the macroscale that can be solved easily and used in predictive and comprehensive studies. But since the model is derived from a microscale model, its parameters bear a direct relation to the microscale features  \citep{hornung1997}. For these reasons, multiscale models have become a popular tool in mathematical modeling \citep[see for example][]{Mikelic2014, ilp2014, Ray15, Schmuck2015, DBG16}.

Let us discuss three studies using the multiscale approach that are the most relevant to the work in this paper. \citet{ilp2014} use a volume averaging approach, which yields the macroscopic equations via local averages in the form of volume integrals. The proposed model accounts for the microscale features of the filtration while modelling a whole filter element, that is, a casing for the filter medium with an inlet for the contaminated fluid and an outlet for the filtered fluid. However, the computational complexity of the resulting model is still quite challenging, requiring resolution of the filter media thickness at the microscale in addition to performing separate simulations for multiple location in the filter media. Thus, while this model is good to understand the effect of the microstructure on the filtration behaviour for a single set of parameters, performing studies for different types of  microstructure using such a model is not practical.

The models by \citet{Ray15} and \citet{DBG16} employ the method of multiple scales, which assumes a separation of scales and averages the microscale variations. Both papers consider the flow and particle transport problems in an evolving porous media, and \citet{Ray15} also accounts for a general interaction potential between fibres and particles (such as an electrostatic potential). Their multiscale models consist of a coupled system of equations for the flow and transport with the effective parameters determined by solving the so-called \emph{cell problems} in a microscopic unit cell. The model by \citet{Ray15} considers a more general microstructure using a level-set framework, but the downside is that the microscopic and macroscopic problems are fully coupled, meaning that the cell-problems have to be solved for each point in space and time. Under certain simplifications, namely no interaction potential and a quasi-periodic microstructure with unidirectional fibres that grow radially due to contaminant deposition, the model by \citet{Ray15} reduces to the one derived by \citet{DBG16}. Under these assumptions, the cell problems depend only on the porosity and so the microscopic and macroscopic problems decouple, resulting in a more efficient simulation. On the other hand, the applicability of the model from \citep{DBG16} is limited due to its consideration only of microstructures of filter media with monodisperse fibres located on a regular lattice. Moreover, under their model assumptions, the simulation must be stopped when two fibres become in contact due the contaminant deposition.

\subsubsection*{Overview} 
In this paper, we use the method of multiple scales to study the effect of the filter microstructure in various filtration regimes. We are concerned with nonwoven filter media, which is one of the most common filter media types \citep[see][]{Brown1993,Hutten2015}. The nonwoven medium is a sheet made from directionally or randomly oriented fibres bonded together by chemical, mechanical, heat or solvent treatment. The contaminants are transported via diffusion and advection with the fluid and are considered to be small enough in comparison with the fibre size to use a continuum approach. 

Our set-up is similar to the one in \citep{DBG16}. In that paper the authors assumed that fibres are arranged in a simple quasi-periodic structure (a hexagonal lattice), and that all fibres have the same radius in a given unit cell. But real filter media have fibres with some diameter distribution, that is, \textit{polydisperse fibres}, and do not have a regular fibre arrangement. To this end, in this paper we allow for different fibre sizes in the same unit cell and for random microstructures. By random microstructure we mean a unit cell with randomly distributed fibres that is representative of the material as a whole and then extended quasi-periodically. Quasi-periodicity means that we allow for slow variations from unit cell to cell to enable us to capture porosity variations on the macroscale (present either initially by design of the filter media or due to nonuniform contaminant adsorption). 

In regular microstructures with equally sized fibres, fibres are grown (due to contaminant deposition) until the close-packing of the given lattice is reached and the simulation is terminated \citep{DBG16}. In a random lattice this method would not work well since two fibres could already be close initially, leading to a short filter lifetime. To deal with this case, we propose an agglomeration algorithm whereby as fibres come into contact they are combined into a larger fibre.

The structure of the paper is as follows. In \cref{sec:model}, we present our model and the algorithm for joining fibres. Then, we perform a comprehensive study on how the microstructure influence the effective parameters of the filter media in \cref{sec:micro}. We consider five different microstructures: regular square, regular hexagonal and three random with different inter-fibre properties. Then, we discuss how the effective parameters are affected by microstructure differences. In \cref{sec:macro_method} we discuss criteria used to evaluate the performance of filter media and different filtration regimes and operational set-ups. In particular, we consider filtration when the contaminant transport occurs due to advection, diffusion or both, and operation set-ups with either constant flow rate or constant pressure drop. In \cref{sec:multi_sim}, we carry out multiscale simulations for the five types of microstructures and different filtration regimes and set-ups. Here, we discuss and investigate in detail how each regime is affected by the filter-media microstructure. We perform further analysis of the transport mechanisms of the contaminants in \cref{sec:transport} and investigate how the initial efficiency is influenced by contribution of the advection and diffusion terms. Finally, in \cref{sec:conc} we summarize our findings.


\section{Mathematical model \label{sec:model}}


In this section we present the derivation of the multiscale model. We consider the general case when the transport of contaminant particles is due to a combination of diffusion and advection in a fluid flow. The cases when transport is only diffusive or advective are contained in this model. 

\begin{figure}
\centering
\begin{minipage}[b]{.48\textwidth}
  \centering
  \includegraphics[width=1\linewidth]{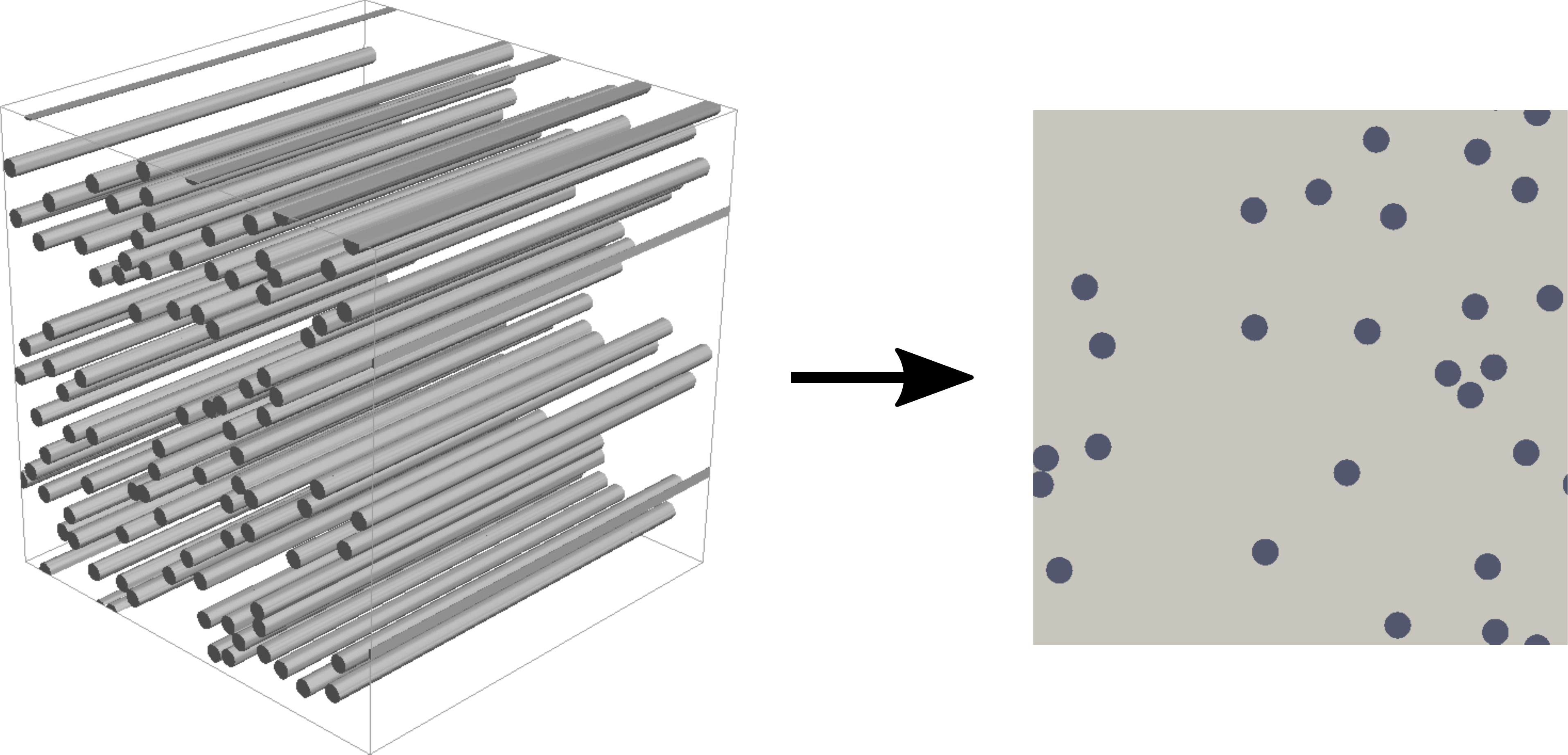}
  \caption{Nonwoven filter media microstructure with unidirectional fibres: 3D and 2D representations {on the left and right}, respectively}
  \label{fig:2dto3d}
\end{minipage}
\hfill
\begin{minipage}[b]{0.48\textwidth}
  \def\svgwidth{1\linewidth} 
  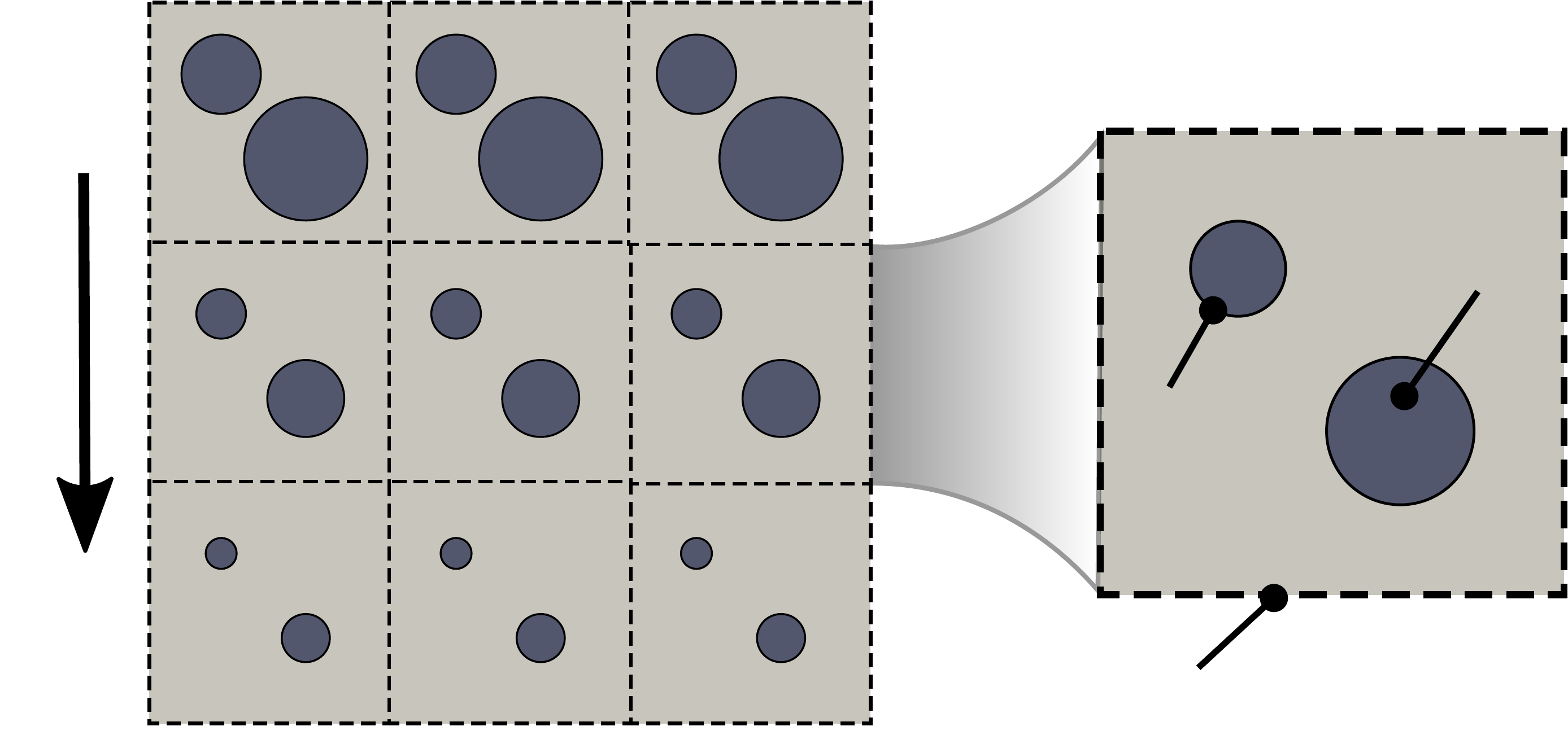
  \caption{Microscale representation of the filter media {on the left} and notations for the microscale quasi-periodic unit cell $\midmn$ {on the right}}
  \label{fig:midmn}
\end{minipage}
\end{figure}

We begin by describing the problem at the microscopic, or pore, scale, at which we assume the media has initially a known and periodic microstructure that consists of so-called unit cells. Within each unit cell we allow fibres of different sizes to be present. We assume that the fibres grow radially as contaminants adsorb onto their surface. Contaminant adsorption occurs at different rates in different unit cells, depending on the local particle concentration and flow. As a result, fibres in unit cells can grow differently (see the schematic on the left of \cref{fig:midmn}). However, we assume that the variations in diameter of the periodic fibres are small between adjacent unit cells, so that our microstructure is near-periodic and these variations are captured at the macroscale.

 We suppose that the media is composed of unidirectional fibres, which naturally reduces the model to a two-dimensional microstructure (see \cref{fig:2dto3d}), though we note that all of the analysis presented here readily extends to three dimensions. 

The macroscopic domain is denoted $\dmndim\subset\mathbb{R}^2$ and consists of the fluid and solid subdomains $\dmnfdim(\tilde{t})$ and $\dmnsdim(\tilde{t})$, respectively, where tildes denote dimensional quantities. The solid subdomain $\dmnsdim$ represents the fibres. The interface between the subdomains is denoted $\bdrysdim(\tilde{t}) = \dmnfdim\cap\dmnsdim$ and represents the total surface of the fibres.  We note that both subdomains depend on time $\tilde{t}$ due to fibre growth. 


\subsection{Microstructure with isolated fibres \label{sec:model1}}

First, we consider the case when fibres are not allowed to touch at any time $\tilde t$. We recall this was the assumption used by \citet{DBG16}, but here we extend it to polydisperse fibres. We define the domain by setting the location of the fibres in $\dmndim(\tilde t)$ with centres located at $\fibcentdim$ and radii $\miradiusfibdim(\tilde{\x},\tilde{t})$ for $\allfib$, where $\fibset$ is the set of all fibres in the whole filter media, and prescribe  $\dmndim(\tilde t = 0)$. We denote the surface of each fibre as:
\begin{equation}
  \bdrysfibdim(\tilde t) = \left\{\tilde{\x}\in\dmndim: 
  \norm{\tilde{\x}-\fibcentdim} 
  = \miradiusfibdim(\fibcentdim,\tilde{t})\right\}, \:\allfib.
\end{equation}
The the interface between the pore and solid subdomains is the union of all the fibre surfaces, that is, $\bdrysdim(\tilde{t}) = \bigcup_{\allfib}\bdrysfibdim(\tilde{t})$. We note that the assumption that fibres are not in contact implies that $\bdrysdim^m(\tilde{t}) \bigcap \bdrysdim^n (\tilde t) = \emptyset$ for $n \neq m$. 

The contaminant particles and the fluid occupy the pore space of the filter media, $\dmnfdim(\tilde{t})$. We assume that the particles are sufficiently small that they do not influence the fluid flow and that the flow is incompressible and Newtonian. We also assume that the flow is sufficiently slow and thus satisfies the Stokes equations:
\begin{subequations}
\label{dimstokes}
\begin{alignat}{2}
  \label{dimstokes1}
  -\tilde{\nabla}\tilde{\mipres}+\visc\tilde{\nabla}^2\tilde{\mivel} 
  &= \mathbf{0}, \qquad
  &\tilde{\x}&\in\dmnfdim(\tilde{t});\\
  \label{dimstokes2}
  \tilde{\nabla}\cdot\tilde{\mivel} &= 0, 
  &\tilde{\x}&\in\dmnfdim(\tilde{t});
\end{alignat}
where 
$\tilde{\mipres}(\tilde{\x},\tilde{t})$ is the fluid pressure in $[Pa]$, $\tilde{\mivel}(\tilde{\x},\tilde{t})$ is the fluid velocity in $[m/s]$, $\tilde{\nabla}$ is the nabla operator with respect to the spacial coordinate $\tilde{\x}$, and $\visc$ is the viscosity in $[Pa\,s]$. The radial fibre growth results in the following no-slip boundary condition on the fibre surface:
\begin{alignat}{2}
  \label{dimstokes3}
  \tilde{\mivel} 
  &= -\partDeriv{\miradiusfibdim}{\tilde{t}}\normal^{\fib},\qquad
  \tilde{\x}\in\bdrysfibdim(\tilde{t}), \:\allfib;
\end{alignat}
\end{subequations}
where $\normal^{\fib}$ is the unit normal to the fibres' surface $\bdrysfibdim$ pointing into the solid domain. Despite the time-dependent nature of the boundary condition, \cref{dimstokes3}, we use the steady-state Stokes \cref{dimstokes1,dimstokes2} because the timescale of fibre growth is slow compared with that over which the fluid flow attains a steady state.

We assume that the contaminant particles are uniform in size and, as is common in filtration scenarios, much smaller than the typical fibre diameter. This allows us to consider the contaminant as a continuum and describe it by its number concentration $\tilde c(\tilde{\x},\tilde{t})$, which evolves according to:
\begin{equation}
  \label{dimmass}
  \partDeriv{\miconcdim}{\tilde{t}} 
  = \tilde{\nabla}\cdot\lr{\diff\tilde{\nabla}\miconcdim
  - \tilde{\mivel}\miconcdim},
  \quad \tilde{\x}\in\dmnfdim(\tilde{t});
\end{equation}
where $\miconcdim$ is measured in $[\textit{particle count}/m^3]$, and $\diff$ ($[m^2/s]$) is the diffusivity coefficient. Using a linear adsorption model and \cref{dimstokes3}, the boundary condition for the concentration reads
\begin{equation}
  \label{dimcbc}
  -\diff\tilde{\nabla}\miconcdim\cdot\normal^{\fib} = \adsdim\miconcdim,
  \quad \tilde{\x}\in\bdrysfibdim(\tilde{t}), \:\allfib;
\end{equation}
where $\adsdim$ ($[m/s]$) is the adsorption coefficient. This represents a balance between the diffusive flux of particles to the fibre surface and the net adsorption as a result of contact, which, as discussed in the Introduction, may be due to a combination of mechanisms.

We assume that the contaminant particles become immobile once they adsorb onto the fibre surface  and add to the fibre volume. Since we assume that the fibres grow radially, the radius of each fibre changes proportionally to the volumetric particle flux averaged over the surface of the fibre $\bdrysfibdim$. This implies:
\begin{equation}
  \label{dimcoupling}
  \partDeriv{\miradiusfibdim}{\tilde{t}} 
  = \frac{1}{|\bdrysfibdim|} \int_{\bdrysfibdim}
  \pack^{-1}\partvol\,\adsdim\miconcdim \dint s,\qquad
  \tilde{\x}\in\bdrysfibdim(\tilde{t}), \:\allfib,
\end{equation}
where $|\bdrysfibdim|=\int_{\bdrysfibdim}\dint s$, $\partvol$ ($[m^3]$) is the volume of a contaminant particle, and $\pack$ is the packing density of contaminant particles on the fibre surface. If we ignored voids between the contaminant particles adsorbed onto the fibre surface, then $\pack=1$; if they are perfectly packed around the fibre, then $\pack = 0.74$. Generally, the particles do not pack so well and form so-called dendrites, or very dense tree-like structures \citep[pp.201--205]{Brown1993}. We consider $\pack=0.3$ in this study. 

\subsubsection{Nondimensionalization}

We introduce the following nondimensionalization: 
\begin{equation}
\label{dimlessvar}
\begin{aligned}
\tilde{\x} = l\x,\quad
\tilde{\mivel} = \mathcal{U}\mivel,\quad
\tilde{\mipres} = \frac{\visc \mathcal{U}}{\delta^2 l}\mipres,\quad
\tilde{t} = \mathcal{T} t,\quad
\miradiusfibdim = \delta l \miradiusfib,\quad
\miconcdim = \conctot \miconc,
\end{aligned}
\end{equation}
where $l$ is the characteristic thickness of the filter media in $[m]$, $\mathcal{U}$ is the characteristic face velocity in $[m/s]$, $\mathcal{T}$ is the characteristic filtration time in $[s]$, and $\conctot$ is the inlet contaminant concentration in  $[\textit{particle count}/m^3]$. Here $\delta$ is the ratio of the microscopic quasi-periodic unit cell diameter to the (macroscopic) filter depth. The assumption to apply the method of multiple scales is that there is a separation of between the microscopic and macroscopic lengthscales, that is, $\delta \ll 1$. 

We also introduce the following dimensionless groups:
\begin{equation}
  \label{dimconst}
  \coefA = \frac{l}{\mathcal{U}\mathcal{T}}, \quad
  \coefB = \frac{l\delta}{\ads\mathcal{T}}, \quad
  \coefC = \frac{\diff\delta}{l\ads}, \quad
  \coefD = \frac{\mathcal{U}\delta}{\ads}, \quad
  \coefE = \frac{\mathcal{T}\partvol\mathcal{C}\ads}{\pack\delta l}
\end{equation}
and the dimensionless domains $\dmn$, $\dmnf$, $\bdrys^{\fib}$, analogous to their dimensional counterparts. Inserting this nondimensionalization into \cref{dimstokes} yields:
\begin{subequations}
\label[equation]{stokes}
\begin{alignat}{2}
  \label{stokes1}
  -\nabla\mipres+\delta^2\nabla^2\mivel &= \mathbf{0}, 
  &\x&\in\dmnf(t);\\
  \label{stokes2}
  \nabla\cdot\mivel &= 0, 
  &\x&\in\dmnf(t);\\
  \label{stokes3}
  \mivel &= -\delta\coefA\partDeriv{\miradiusfib}{t}\normal^{\fib},\qquad
  &\x&\in\bdrysfib(t),\:\allfib.
\end{alignat}
\end{subequations}
The mass-transport problem, \cref{dimmass,dimcbc}, transforms into:
\begin{subequations}
\label[equation]{mass}
\begin{alignat}{2}
  \coefB\partDeriv{\miconc}{t} 
  = \nabla\cdot\lr{\coefC\nabla\miconc
  - \coefD\mivel\miconc}&,
  &\x&\in\dmn_f(t);\\
  \coefC\nabla\miconc\cdot\normal = -\delta \miconc&,\qquad
  &\x&\in\bdrys(t).
\end{alignat}
\end{subequations}
Finally, using \cref{dimcbc,dimcoupling}, the coupling condition in dimensionless form reads:
\begin{equation}
  \label{coupling}
  \partDeriv{\miradiusfib}{t} = 
  \frac{1}{\abs{\bdrysfib}}
  \int_{\bdrysfib}
  \coefE\miconc
  \dint s,\:\allfib.
\end{equation}


\subsubsection{Homogenized model \label{sec:homog}}

At the microscale, we consider the filter media to consist of quasi-periodic unit cells $\miDmn$ (see \cref{fig:midmn}). We allow for fibres of different sizes that may be randomly arranged within each periodic cell. We introduce a microscale variable $\y = \x/\delta$, which is defined in the unit cell $\miDmn$. The macroscale variable $\x$ spans across the whole filter media. We denote the fluid and solid subdomains as $\miDmnF$ and $\miDmnS$, respectively. The internal fluid--solid interface is denoted as $\miBdryS$, which consists of the fibre surfaces $\miBdryS=\cup_{\fib\in\mifibset}\mibdrysfib$, where $\mifibset$ is the set of the fibres found in the unit cell, which is a subset of all fibres $\fibset$ in the filter media, $\mifibset\subset\fibset$. The outer fluid boundary of the unit cell is denoted $\miBdryF = \mibdry\cap\midmnf$.

Using the method of multiple scales, we seek a solution to the problem \cref{stokes,mass,coupling} as a function of $\x$ and $\y$, and treat these two variables as independent. The extra freedom this gives is removed by enforcing that the solution is exactly periodic in $\y$; small variations from one unit cell to the next are thereby captured through the macroscale variable $\x$. We insert the change of variables $\y = \x/\delta$ and expand all dependent variables in the form $\mivel = \mivel^{(0)} + \delta \mivel^{(1)} + \cdots$ and similarly for the pressure $\mipres$ and the concentration $\miconc$.

The macroscopic quantities are introduced using the following averaging:
\begin{equation}
\label{avg}
  \overline{G}(\x,t,\cdot) 
  = \frac{1}{\abs{\miDmn}}\int_{\midmnf}g(\x,\y,t,\cdot)\dint\y 
  = \phi G(\x,t,\cdot),
\end{equation}
where $g(\x,\y,t)$ is a microscopic quantity, $\overline{G}(\x,t)$ is its volumetric average, $G(\x,t)$ is its intrinsic average, and $\phi = \abs{\midmnf}/\abs{\midmn}$ is the porosity.

For our homogenization, we use the volumetric average for the velocity $\velv$ and the intrinsic average for the pressure $\pres$ and concentration $\conc$. The velocity $\velv$ is the Darcy velocity and should not be confused with the actual velocity of the fluid travelling through the pores. 

The derivation of the homogenized model via the method of multiple scales is analogous to that presented by \citet{DBG16} for monodisperse fibres. The same methodology can be easily extended to fibres with polydisperse sizes in the unit cell, when the fibres do not touch. For this reason, here we just present the final homogenized model and refer the reader to \citet{DBG16} for the details. In what follows, we drop the superscript $(0)$ that refers to the leading-order quantities in $\delta$ and simply write $\velv^{(0)} \equiv \velv$, and similarly for $\pres$ and $\conc$.


The homogenization of the flow problem \cref{stokes} leads to Darcy's equation, which relates $\velv$ and $\pres$ in terms of the permeability tensor $\perm$:
\begin{subequations}
\label[equation]{flow1}
\begin{align}
  \label{flow1a}
  \velv(\x,t) &= -\perm\nabla_{\x}\pres,\\
  \label{flow1b}
  \perm(\x,t) &= \frac{1}{\abs{\midmn}}\int_{\midmnf}\mathbf{K}\dint\y.
\end{align}
\end{subequations}
Here, $\mathbf{K}$ is a matrix-valued function and together with a vector-valued function $\mathbf{\Pi}$ they satisfy the following cell problem at each location $\x$:
\begin{subequations}
\label[equation]{cell_flow}
\begin{alignat}{2}
  \mathbf{I} - \nabla_{\y}\mathbf{\Pi} + \nabla_{\y}^2\mathbf{K} &= \mathbf{0},
  &\y&\in\miDmnF;\\ 
  \nabla_{\y}\cdot\mathbf{K} &= 0,
  &\y&\in\miDmnF;\\ 
  \label{cell_flow_c}
  \mathbf{K} &= \mathbf{0},
  &\y&\in\miBdryS;\\
  \mathbf{K}&,\,\mathbf{\Pi} \text{ periodic},\qquad
  &\y&\in\miBdryF;
\end{alignat}
\end{subequations}
where $\mathbf{I}$ is the identity matrix. For isotropic filter media, the permeability tensor becomes a multiple of the identity, that is, $\perm =\sPerm  \mathbf{I}$ with $\sPerm$ being a scalar.
The macroscopic analogue to the incompressibility condition, \cref{stokes2} is:
\begin{equation}
  \label{flow2}
  \nabla_{\x} \cdot \velv 
  = \frac{\alpha}{\abs{\midmn}}\sum_{\allmifib}\int_{\mibdrysfib}
  \partDeriv{\miradiusfib}{t} \dint s
  = \frac{\alpha}{\abs{\midmn}}\sum_{\allmifib}\partDeriv{\miradiusfib}{t}(2\pi\miradiusfib)
  = -\alpha\partDeriv{\phi}{t},
\end{equation}
where $\alpha$ is given in \cref{dimconst}. 


Under the multiple-scales transformation, the contaminant-transport equation \eqref{mass} becomes:
\begin{equation}
  \label{ma_mass}
  \coefB\partDeriv{\lr{\phi\conc}}{t}
  =
  \nabla_{\x}\cdot\lr{\coefC\phi\deff\nabla_{\x}\conc - \coefD \velv \conc}
  - \area\conc,
\end{equation}
where $\beta$ and $\gamma$ are given in \cref{dimconst} and $\area$ is the effective surface area of the fibres defined as $\area=\abs{\mibdrys}/\abs{\midmn}$ (that is, the surface area per unit volume of media). 
The effective diffusion coefficient $\deff=\deff(\x,t,\partd)$ is computed as:
\begin{equation}
  \label{deff}
  \deff = \mathbf{I} - \frac{1}{\abs{\midmnf}}\int_{\midmnf}\mathbf{J}^T_{\mathbf{\Gamma}}\dint\y,
\end{equation}
where $\lr{\mathbf{J}^T_{\mathbf{\Gamma}}}_{ij} = \partial\Gamma_j/\partial y_i$ is the transpose of the Jacobian matrix of the vector-valued function $\mathbf{\Gamma}$. Its components $\Gamma_i$ satisfy the cell problem:
\begin{subequations}
\label[equation]{mass_cell}
\begin{alignat}{2}
  \nabla_{\y}^2\Gamma_i &= 0,
  &\y&\in\miDmnF;\\ 
  \nabla_{\y}\Gamma_i \cdot \normal^{\fib}_{\y} &= \lr{\normal^{\fib}_{\y}}_i,
  &\y&\in\miBdrySFib,\:\allmifib;\\
  \Gamma_i &\text{ periodic}, \qquad
  &\y&\in\miBdryF;
\end{alignat}
\end{subequations}
where $\lr{\normal^{\fib}_{\y}}_i$ are the components of $\normal^{\fib}_{\y}$. We note that in an analogous fashion to the permeability, we have $\deff=\sDeff \mathbf{I}$ for the case of isotropic filter media.


Finally, we obtain the following macroscopic coupling condition from \cref{coupling}:
\begin{equation}
\label{eq:drdt2}
  \partDeriv{\miradiusfib}{t}
  =
  \coefE\conc,\:\allmifib,
\end{equation}
where $\coefE$ is given in \cref{dimconst}. Multiplying \cref{eq:drdt2} by $\abs{\mibdrysfib}/\abs{\midmn}$ and summing over all fibres in the unit cell, we obtain the following relation between the fibre radii and the macroscopic porosity:
\begin{equation} \label{eq:r_porosity}
  \sum_{\allmifib}\frac{\abs{\mibdrysfib}}{\abs{\midmn}}\partDeriv{\miradiusfib}{t} 
  = 
  \frac{1}{\abs{\midmn}}\partDeriv{}{t} \lr{\sum_{\allmifib}\pi\lr{\miradiusfib}^2}
  =  -\partDeriv{\phi}{t}.
\end{equation}
Combining \cref{eq:drdt2,eq:r_porosity} yields:
\begin{equation}
  \label{ma_coupling}
  \partDeriv{\phi}{t} = -\coefE\area\conc.
\end{equation}

The diffusion of contaminant particles has a two-fold impact on the filtration process. First, it acts as a bulk transport mechanism of the contaminant particles, appearing in the dimensionless parameter $\coefC$. Second, it corresponds to the driving feature in the capture mechanism, expressed by \cref{ma_coupling}. The molecular diffusion $\diff$ and correspondingly the dimensionless parameter $\coefC$ scale as $q^{-1}$, where $q$ is the size of the contaminant particles, while the diffusion component in the adsorption coefficient $\ads$ scales as $q^{-2/3}$ according to an empirical adsorption model from \citet[pp. 205--210]{Baron2001}. This means that as particle size increases the molecular diffusion~$\diff$ converges to zero faster than the diffusion effect in the adsorption. As a result, it is possible to have a filtration regime where the diffusion term in \cref{ma_mass} is negligible but particle adsorption is still mostly driven by diffusion. This will be the case in our \emph{advection only} regime. 

For the sake of clarity, in this study we shall assume that $\coefE$ is constant for all filtration regimes and for all fibres, but recognize that in reality it might differ for each fibre and may also change with time. If we wanted to allow for different adsorption coefficients for different fibre surfaces, we would need to derive the equations in terms of an effective adsorption coefficient instead of the effective surface area~$\area$, but we do not consider this here.


\subsection{Microstructure with closely located fibres \label{sec:join}}

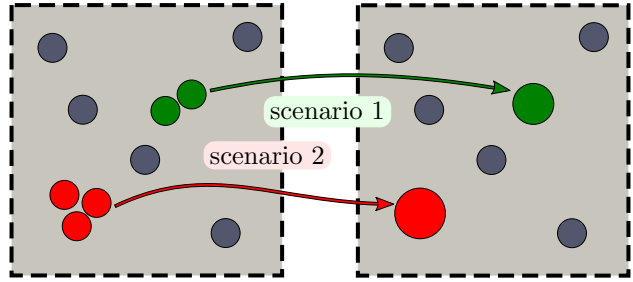
\begin{figure}
    \centering
	\begin{tikzpicture}
    \node[inner sep=0] at (0,0) {\includegraphics[width=0.5\linewidth]{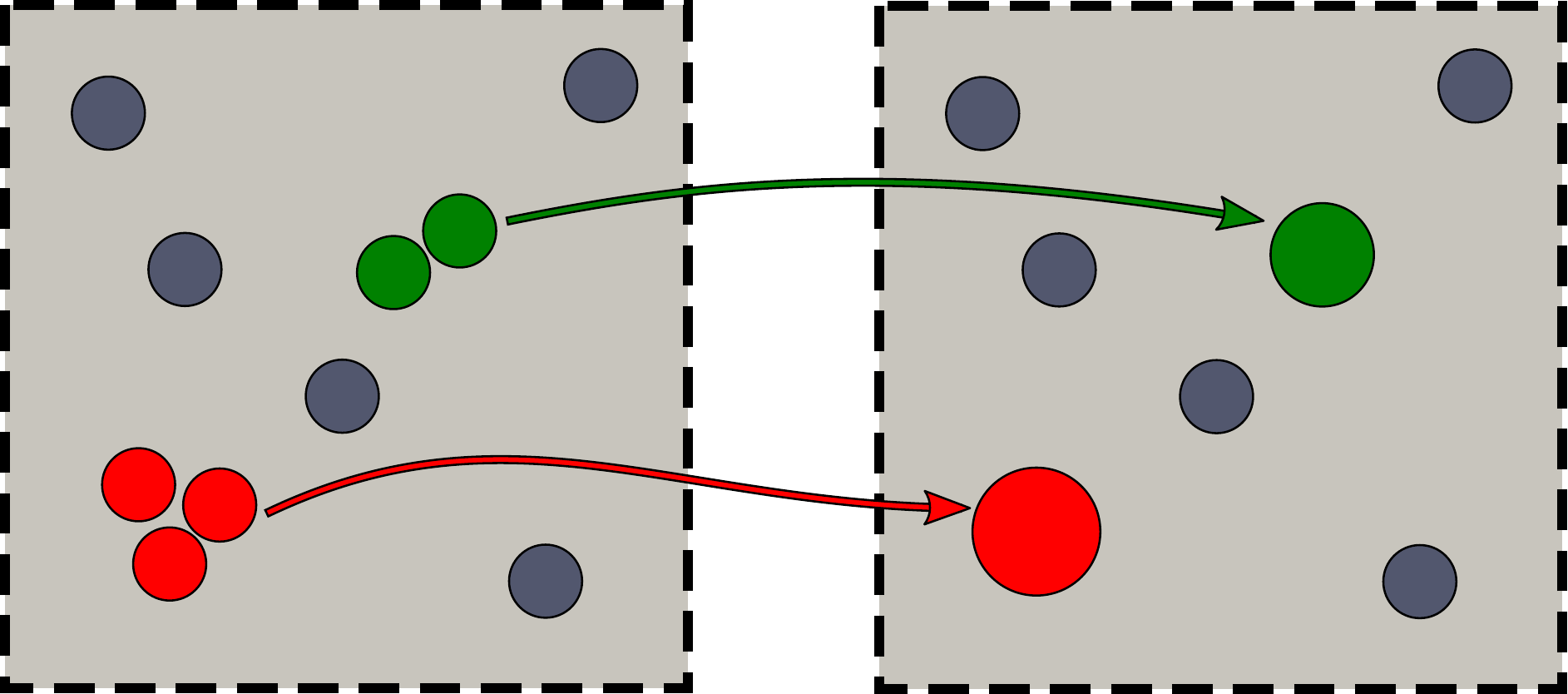}};
    \draw[fill=green!10, green!10, rounded corners] (-0.74, 0.58) rectangle (0.94, 0.18);
    \draw[fill=red!10, red!10, rounded corners] (-1.54, -0.02) rectangle (0.14, -0.38);
    \node at (0.1, 0.4) {scenario 1};
    \node at (-0.7, -0.2) {scenario 2};
	\end{tikzpicture}
    \caption{Agglomeration algorithm of touching fibres. Scenario 1: two fibres come into contact with one another at some point and are unified into a single fibre with the same volume with the same centre of mass. Scenario 2: two fibres come into contact with one another at some point and are unified into a single fibre which then overlaps with a neighbouring fibre. The unified fibre is then joined with the third fibre to form a fibre with the same volume and centre of mass as the three fibres.}
    \label{fig:joining}
\end{figure}

In random microstructures, the distance between different fibres varies (see \cref{fig:2dto3d}). As contaminants are being deposited and fibres are growing, fibres located close to one another come into contact and form an agglomerate, while other more distant fibres can continue growing individually. To account for this scenario, we introduce the following agglomeration algorithm. If, as fibres grow radially, two or more fibres come into contact, we replace them with one larger fibre located at the centre of mass of the original fibres and with cross-section area equal to the sum of the areas of the individual fibres (see scenario 1 in \cref{fig:joining} for an illustrative example). When replacing two fibres with a single larger fibre, the resulting fibre may overlap with other fibres located nearby in the unit cell. In such an instance, we recursively replace the overlapping or closely located fibres until the resulting fibre becomes isolated (see scenario 2 in \cref{fig:joining}). Due to the periodic boundary conditions on the unit cell, the agglomeration algorithm also accounts for the periodic images of fibres. 

The multiscale model of \cref{flow1,cell_flow,flow2,ma_mass,mass_cell,deff,ma_coupling} is valid for any random and polydisperse configuration of fibres with different radii in the same cell unit. When, due to the fibre growth \cref{eq:drdt2}, two or more fibres come into contact, we perform the geometry transformation described above and then re-apply the same model to the new geometry. 

While this algorithm is clearly an idealization of the real process, it allows us to account for the formation of fibre agglomerates while keeping our strategy simple and robust. In \cref{app_1} we investigate the effect of the fibres coming into contact on the numerical simulations, in particular to confirm that the effective diffusivity~$\deff$ converges to a limiting case.

The continuum assumption that was used to model the contaminant transport is violated as the distance between two fibres become comparable with the particle size. To resolve this issue one can introduce into the agglomeration algorithm a critical distance between two fibres at which they coalesce to form an agglomerate, but we do not include this in our analysis here. 


\subsection{Multiscale algorithm}

In this subsection, we describe the numerical implementation of the multiscale model. First, we perform the microscale simulations as a preprocessing step and find the effective parameters, namely, the permeability $\perm$, the effective diffusivity $\deff$ and the effective surface area $\area$, as functions of the porosity $\phi$. The description of the microscale simulations is presented as a schematic algorithm in \cref{fig:alg}. As input parameters we specify the microstructure type, and the initial porosity and fibre diameter distribution. Using this input, we generate one unit cell in the case of regular microstructures (square or hexagonal), and multiple random instances of unit cells in the case of random microstructures. Once the unit cell for the given porosity is characterized (with the parameters of interest listed above), we decrease the porosity by a small porosity step $\Delta \phi$. Then, we increase the fibre radii until the new porosity is reached and compute the effective parameters corresponding to the updated porosity value. If necessary, we apply the agglomeration algorithm described in \cref{sec:join}. We continue this process until we have reconstructed the whole dependency of the effective parameters on the porosity (see for example \cref{fig:micro1,fig:micro2}). We note that if we were to take a microstructure configuration obtained at a later time from this algorithm and run the process in reverse (increasing rather than decreasing porosity) then we would not recover the earlier configurations since we lose information about the original microstructure upon merging fibres. We could however, create a different algorithm to describe a scenario in which the obstacles decrease in size and divide. 

We implement the microscopic algorithm in Python. We generate the unit cells and a triangular unstructured grid with refinement around the fibre surfaces using the open-source mesh generator GMSH \citep{gmsh}. Then, we discretize the cell problems using the finite element library FEniCS \citep{fenics} using linear Lagrange elements to solve \cref{mass_cell} and a mixed finite element method to solve \cref{cell_flow} with linear and quadratic Lagrange elements for $\mathbf{\Pi}$ and $\mathbf{K}$, respectively. 

\begin{figure}
\begin{algorithm}[H]
\caption{Microscale simulations}
\SetKwFunction{CreateGeom}{CreateGeometry}
\SetKwFunction{CompParams}{ComputeEffectiveParameter}
\SetKwFunction{FindRadius}{GrowFibres}
\SetKwFunction{CompAvg}{ComputeMean}
\SetKwFunction{CompErr}{ComputeMonteCarloError}
 \KwData{porosity $\phi\leftarrow\phi(t=0)$, fibre radius $\miradius^{\fib}\leftarrow\miradius^{\fib}(t=0)$ for $\fib\in\mifibset$}
 \KwResult{effective parameters $\perm$, $\deff$, $\area$ as functions of porosity $\phi$}
 \While{possible to create a unit cell with $\phi$ and $\miradius^{\fib}$}{
 \For{$\mathcal{P}\in\{\perm, \deff, \area\}$}{
  \eIf{microstructure is random}{
    $l \leftarrow 1$\;
    $\mathcal{P}^{MC}\leftarrow 0$\;
    \While{Monte Carlo error $>$ accuracy}{
    $\midmn\leftarrow$ \CreateGeom{$\phi$, $\miradius^{\fib}$}\;
    $\mathcal{P}\leftarrow$ \CompParams{$\midmn$}\;
    $\mathcal{P}^{MC}\leftarrow$ \CompAvg{$\mathcal{P}^{MC}$, $\mathcal{P}$}\;
   }
   $\mathcal{P}\leftarrow\mathcal{P}^{MC}$
   }{
    $\midmn\leftarrow$ \CreateGeom{$\phi$, $\miradius^{\fib}$}\;
    $\mathcal{P}\leftarrow$ \CompParams{$\midmn$}\;
   }
 }
 $\phi\leftarrow\phi-\Delta\phi$\;
 $\miradius^{\fib}\leftarrow$ \FindRadius{$\phi$, $\miradius^{\fib}$}\;
 }
\end{algorithm}
\caption{Schematic algorithm for microscale simulations.}
\label{fig:alg}
\end{figure}

Once we know the dependencies of the effective parameters on the porosity, we use them as coefficients in the macroscale model \cref{flow1,flow2,ma_mass,ma_coupling}. The system of macroscale equations is also implemented in Python, discretizing time with an implicit backward Euler scheme and space finite elements from FEniCS library using quadratic Lagrange elements for the pressure and linear Lagrange elements for all other variables. Starting with an initial guess, we find the solution iteratively in time. For a given time step, we solve the mass conservation equation for the contaminant, ~\cref{ma_mass}, and the equation for the contaminant adsorption,~\cref{ma_coupling}, as a fully coupled nonlinear system using the Newton method. Then, we find the corresponding pressure and velocity distributions from~\cref{flow1,flow2} using the obtained porosity and proceed to the next time step. 


\section{Microscale simulations \label{sec:micro}}


In this section, we apply the microscale part of our homogenized model to quantify the three effective characteristics of different types of filter media: the permeability $\perm$, the diffusivity $\deff$, and the surface area $\area$. In particular, we consider five different microstructures, namely two regular and three random arrangements of fibres. The random microstructures are then extended periodically. 

We model a nonwoven filter media with an initial porosity $\phi(0)=0.93$, which is a typical value for nonwoven filter media used in air filters and purifiers \citep[see, for example, Table~1 in][]{Das2009}. To simulate the contaminant deposition, we employ the agglomeration algorithm discussed in \cref{sec:join} to decrease the porosity $\phi$ and to reconstruct the dependence of the effective parameters on the porosity $\phi$.

As discussed in \cref{sec:model}, we model nonwoven filter media using a unidirectional fibre arrangement, which enables a 2D representation of the microstructure (see \cref{fig:2dto3d}). The choice of this representation was motivated by some preliminary validation carried out using experimental data and full 3D simulations using the commercial software package GeoDict \citep{geodict}. The analysis showed that a 2D representation approximates well the permeability and the effective diffusivity of nonwoven media with high porosity, while the effective surface area $\area$ is the same in the 2D case with unidirectional fibres and the 3D case with random orientation of fibres.

\begin{figure}
  \centering
  \def\svgwidth{0.9\linewidth} 
  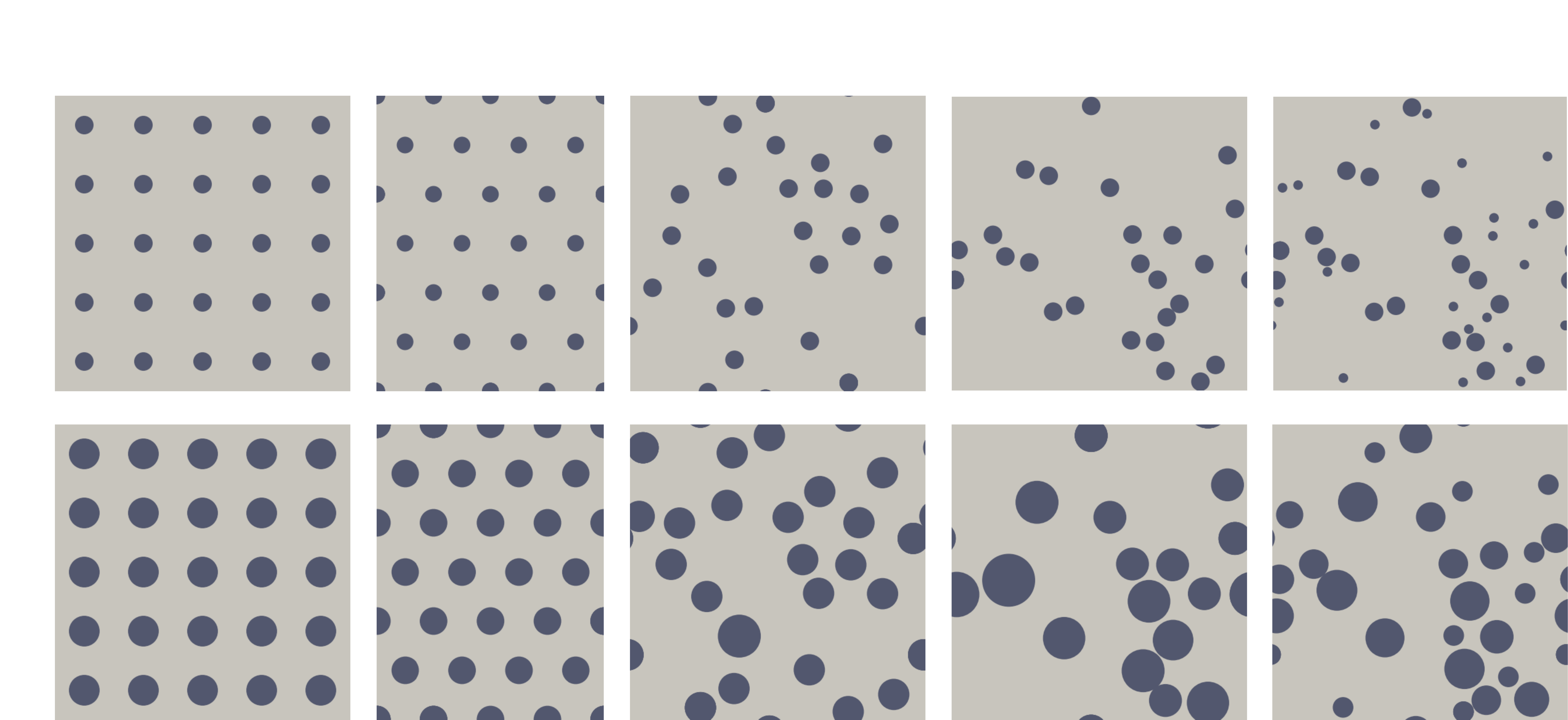
  \caption{Different types of microstructures, that is, square, hexagonal, random with mean isolation distances $2r$ and $0$ and random with polydisperse fibre radii (\textit{from the left to the right}). \textit{The top row} shows microstructures when the filter media is clean with the porosity $\phi(0)=0.93$. \textit{The bottom row} shows the same microstructures after the simulations of the fibre growth due to the contaminant adsorption are performed and the porosity $\phi=0.8$ is reached.}
  \label{fig:microstr}
\end{figure}

The first row of \cref{fig:microstr} shows the five different microstructures corresponding to the clean filter media (before the contaminant deposition has begun). We note that, for the random microstructures, we show one random instance of the fibre distribution but that these will change from sample to sample.  
All microstructures have unidirectional fibres and differ by the distribution of the fibre centres and by the distribution of the fibre radii. The microstructures shown in columns~1--4 in \cref{fig:microstr} have fibres with monodisperse radii at $t=0$, which are denoted $r^{\fib}(\x,0)= r$ constant for $\allfib$, $\x\in\dmn$. The microstructure in \textit{the column~5} has polydisperse fibre radii and will be discussed in detail later. 

We consider regular microstructures with fibre centres located on square and hexagonal grids (see columns~1 and~2 in \cref{fig:microstr}). We then consider three types of random microstructures depending on the fibre position distribution and fibre diameter distribution. In all three cases, fibres are not allowed to overlap (if we try to place a fibre that overlaps, this is rejected and a new position is generated).

The first random microstructure has monodisperse fibre radii and fibre centres uniformly distributed in the unit cell. The problem with this structure is that it can result in areas with many fibres clustered together, and large gaps empty of fibres (see {column~4} in \cref{fig:microstr}). Since this may not be very realistic for some filter media, in the second random microstructure, we impose an additional restriction to have some isolation distance around the fibres  (see {column~3} in \cref{fig:microstr}). While still considering fibres of constant physical radius $r$, the idea is to introduce an average isolation distance so that we cover the unit cell in a more uniform manner. In particular, when placing a new fibre, we set an isolation distance $d_\text{iso} = \max(n, 0)$, where $n \sim \mathcal N(d, d/3)$. Then we attempt to place the new fibre, making sure that the distance between its centre and that of previously placed fibres is at least $2r + d_\text{iso}$ (so note that, once a fibre is placed, we forget about its $d_\text{iso}$). If not, we draw a new candidate position and a new isolation distance for the fibre and try again. The microstructure in {column~3} in \cref{fig:microstr} has the mean isolation distance $d=2r$. The limiting case of the random microstructures with large isolation distance is the hexagonal model, which demonstrates that we can switch between regular and completely random microstructures by changing the isolation distance.

The third random microstructure has uniformly distributed fibre centres with polydisperse fibre radii (see {column~5} in \cref{fig:microstr}). We consider a random microstructure with two different fibre radii: $80\%$ volume-wise of the fibres  have the same as before radius $r$ and the rest $20\%$ are replaced with fibres with radius $r/2$. No isolation distance is imposed in this case, and the initial porosity is preserved to be $\phi(0)=0.93$.

{The second row} in \cref{fig:microstr} shows the microstructures corresponding to the first row after fibres are grown to reach a porosity $\phi=0.8$. The regular microstructures change only the fibre diameter, while in the random ones some agglomerates are formed. Microstructures with large mean isolation distance $d$ ($d=2r$ in  {column~3} in \cref{fig:microstr}) yield fewer agglomerates as contaminants deposited on the fibres in comparison with the small $d$ ($d=0$ in  \textit{column~3} in \cref{fig:microstr}). The microstructure with polydisperse fibre radii also leads to the formation of many agglomerates, but due to the presence of small fibres it preserves a more homogeneous structure during the lifetime of the media than the microstructure with monodisperse fibres and no isolation distance.

\begin{figure}
  \centering
  \includegraphics[width=0.5\linewidth, page=4]{Fig06-08}
  \vskip1mm
  \begin{minipage}[t]{.48\textwidth}
    \centering
    \includegraphics[width=1\linewidth, page=2]{Fig06-08}
    \caption{Permeability $\sPerm$ as a function of porosity $\phi$ for different types of microstructures. The numbers next to the label ``random'' denote the mean isolation distance. For the random microstructures the error bars are not shown, but fall within the markers' size.}
    \label{fig:micro1}
  \end{minipage}
  \hfill
  \begin{minipage}[t]{.48\textwidth}
    \centering
    \includegraphics[width=1\linewidth, page=1]{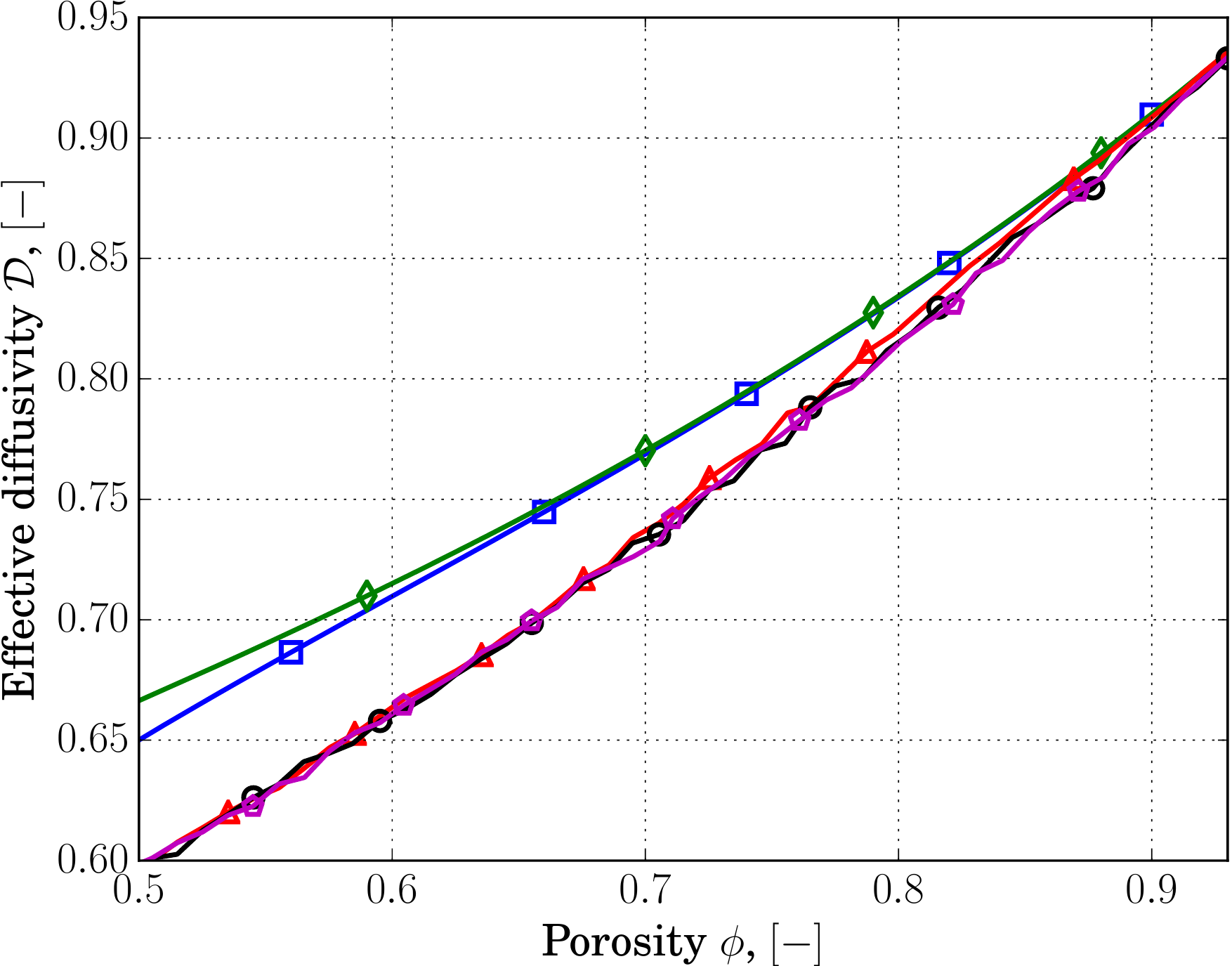}
    \caption{The effective diffusivity $\sDeff$ as a function of porosity $\phi$ for different types of microstructures. The numbers next to the label ``random'' denote the mean isolation distance. For the random microstructures the error bars are not shown, but fall within the markers' size.}
    \label{fig:micro2}
  \end{minipage}
\end{figure}

\begin{figure}
  \centering
  \includegraphics[width=0.5\linewidth, page=4]{Fig06-08}
  \vskip1mm
  \includegraphics[width=0.5\linewidth, page=3]{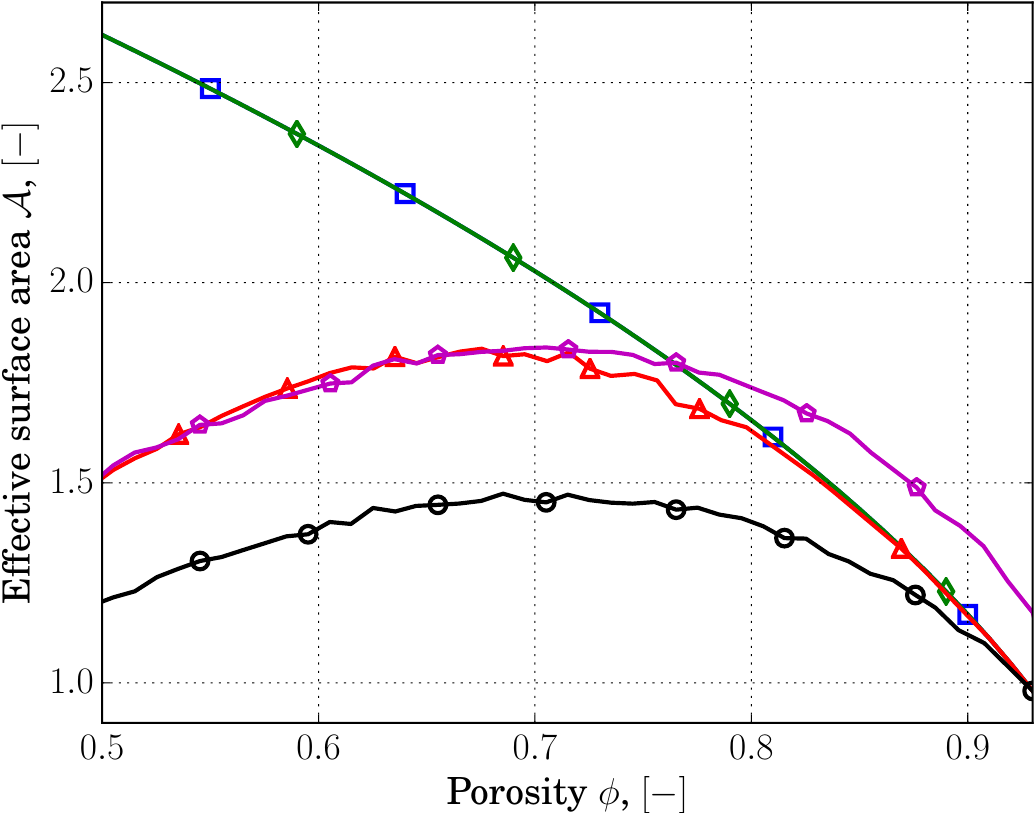}
  \caption{Effective surface area $\area$ as a function of porosity $\phi$ for different types of microstructures. The numbers next to the label ``random'' denote the mean isolation distance. For the random microstructures the error bars are not shown, but fall within the markers' size.}
  \label{fig:micro3}
\end{figure}

We perform microscale simulations for these five microstructure types to obtain the effective parameters, namely the permeability $\perm$, the effective diffusivity $\deff$ and the effective area $\area$, as functions of porosity $\phi$. Since all microstructures considered are isotropic, the permeability and effective diffusivity are presented as scalars, namely $\perm\equiv \sPerm \mathbf{I}$ and $\deff\equiv\sDeff\mathbf{I}$ (see \cref{sec:model1}). For the random microstructures we perform Monte Carlo simulations to find the average effective parameters to an accuracy set to $10^{-2}$. \cref{fig:micro1,fig:micro2,fig:micro3} show the three effective parameters as functions of porosity. The error bars for the random microstructures are not shown, but  they fall within the size of the markers.

The permeability $\sPerm(\phi)$ is computed using \cref{flow1b} after solving the cell problem~\eqref{cell_flow}. We observe that the permeability becomes more sensitive to the microstructure as the porosity decreases (\cref{fig:micro1}). For example, the ratio between the maximum and minimum permeability values for the different microstructures is approximately $1.5$ at $\phi=0.93$ and $5.7$ at $\phi=0.5$. 
The effective diffusivity $\sDeff(\phi)$ is computed using \cref{deff} after solving the cell problem~\eqref{mass_cell}. \Cref{fig:micro2} shows that all random microstructures have the same mean effective diffusivity, while the hexagonal and square microstructures provide slightly higher values. However, overall the effective diffusivity is less sensitive to different fibrous media arrangements than the permeability. Finally, the effective surface area $\area(\phi)$ is the same when the filter media are clean ($\phi=0.93$) for all microstructures except the one with polydisperse fibre radii, but it varies significantly as porosity decreases (\cref{fig:micro3}).

\section{Methodology for macroscale simulations \label{sec:macro_method}}


In this section, we discuss several criteria to measure filter performance and present the operational regimes depending on the contaminant transport mechanism (advection, diffusion, or both) and boundary conditions (filtration at constant pressure drop or constant inflow velocity). 

Our macroscale homogenized model describes the filtration of the contaminants in the porous media, which is characterized by the three effective parameters: permeability, diffusivity and effective surface area. Earlier we assumed that the filter media has unidirectional fibres, which allowed us to describe the filtration problem using a two-dimensional model with a spatial variable $\x=(x,y)$. Now we also assume that the fluid inflow is purely one-dimensional in the direction of the depth of the filter, which we denote by $x$, and periodic boundary conditions in the transverse direction $y$. In addition, we assume the filter media has initially a constant porosity, that is, without macroscopic variations in its microstructure. As contaminant gets deposited in the fibres, the porosity will become a function of the depth $x$ but remain independent of $y$. Then, the macroscopic problem is reduced to one-dimensional with $\x\equiv\sx\in(0,1)$. We denote the scalar velocity by $\svelv$.


\subsection{Filter performance criteria}

The performance of the filter media can be evaluated by different criteria. In this study we consider the following metrics:
\begin{enumerate}
\item Energy consumption: a small pressure drop $\Delta\pres(t) = \pres(0,t) - \pres(1,t)$ across filter media ensures economic use of energy. The pressure drop is determined by the permeability of filter media and the flow velocity.
\item Throughput: a high fluid velocity $\svelv$ achieves large throughput of contaminated fluid across the filter media. We use the fluid velocity at the inlet $\svelv_{in}(t)=\svelv(0,t)$.
\item Efficiency: to quantify how efficient the filter media at trapping the contaminants, we define the number efficiency $\efficiency$ as
\begin{equation}
  \label{eq:eta}
  \efficiency(t) = 1 - \frac{\concflux(1,t)}{\concflux(0,t)},
  \quad\text{ with }\quad\concflux(x,t)=-\coefC \phi\sDeff \frac{\mathrm{d}\conc}{\mathrm{d}\sx} + \coefD \svelv\conc,
\end{equation}
where $\concflux$ is the concentration flux. 
\item Dirt-holding capacity: this tells us how much contaminant the filter media stores over time and, therefore, how long it can be used for. The dirt-holding capacity~$\dhc$ is defined as follows:
\begin{equation}
  \label{eq:dhc}
  \dhc(t) = \pack\int_0^1 \phi(x, 0) - \phi(x,t) \dint x.
\end{equation}
\item Lifetime $T>0$: different termination criteria of the filtration process can be used. For example, in filtration with the constant flow rate, the filter media can be considered completely loaded when a critical pressure drop is reached. On the other hand, in filtration under a constant pressure drop, the lifetime can be determined when the flow rate becomes too small. We choose a criterion that suits both filtration regimes, namely, we set the lifetime $T$ as the time when a minimum porosity $\phi_{min}=0.5$ is reached at any location of the filter media.
\end{enumerate}

To quantify the influence of the microstructure on the filtration performance, we introduce some sensitivity characteristics based on the performance criteria discussed above. First, we consider how sensitive the dirt-holding capacity $H$ is to the microstructure. We compute the relative maximum difference in the dirt-holding capacity while all filter media are in operation:
\begin{equation}
  \dhcsens = \frac{\max_{i,j=1\ldots N}\left|\dhc_i(T^*) - \dhc_j(T^*)\right|}{\max_{i=1\ldots N}\dhc_i(T^*)}, 
  \text{ where }T^*=\min_{i=1\ldots N}T_i;
\end{equation}
where the subscripts denote characteristics corresponding to different microstructures and $N$ is the number of microstructures considered ($N=5$ for us). Second, we quantify the deviations of the pressure drop and the fluid velocity depending on the microstructure:
\begin{equation}
  \{\pressens,\velsens\} = 1 - \frac{\min_{i=1\ldots N} \left|f_i(T_i)-f_i(0)\right|}{\max_{i=1\ldots N} \left|f_i(T_i)-f_i(0)\right|},
  \text{ where }f = 
  \begin{cases}
    \Delta\pres\,\text{ for }\pressens,\\ 
    \svelv_{in}\,\text{ for }\velsens.
  \end{cases}
\end{equation}
Finally, we introduce a sensitivity characteristic for the lifetime of the filter media in the same way we introduced it for the pressure drop deviation:
\begin{equation}
  \timesens = 1 - \frac{\min_{i=1\ldots N} T_i}{\max_{i=1\ldots N} T_i}.
\end{equation}
We note that the sensitivity characteristics $\dhcsens$, $\pressens$, $\velsens$ and $\timesens$ take values in $[0,1]$. When they are close to zero, the respective performance criteria, that is, the dirt-holding capacity, the pressure drop, the the fluid velocity or the lifetime, is not sensitive to the microstructures. In contrast, a sensitivity characteristic close to one implies there is an important difference in the performance criteria between different microstructures.


\subsection{Advection--diffusion regime \label{sec:diffadv}}
\Cref{ma_mass} describes the transport of contaminant by advection and diffusion. Typically for filtration regimes with advection, the timescale for the trapping of contaminant particles is much longer than the timescale for fluid to be advected through the filter media. Therefore, $\coefB\ll1$ and the time derivative in the mass transport  \cref{ma_mass} can be neglected.  Hence, we consider the steady-state of \cref{ma_mass} coupled with the contaminant adsorption  \cref{ma_coupling}. We supplement \cref{ma_mass} with the following boundary conditions. At the inflow boundary $\sx=0$, we specify the contaminant flux $J_{in} = 1$. At the outflow boundary $\sx=1$ we use zero Neumann boundary condition for the concentration. Mathematically this reads \citep[see][]{DBG15}:
\begin{subequations}
\label[equation]{vin_pe_small:bc}
   \begin{alignat}{2}
   \label{vin_pe_small:bc1}
   \left.\lr{-\coefC \phi\sDeff\frac{\mathrm{d}\conc}{\mathrm{d}\sx} + \coefD\svelv\conc}\right|_{x=0} &= J_{in}, \qquad &t\in[0,T];\\
   \label{vin_pe_small:bc2}
   \left.\frac{\mathrm{d}\conc}{\mathrm{d}\sx}\right|_{x=1} &= 0, &t\in[0,T].
   \end{alignat}
\end{subequations}
In addition, we specify an initial condition for the porosity for \cref{ma_coupling}:
\begin{equation}
 \label{vin_pe_small:ic}
 \phi(x,0) = \phi_0, \qquad\sx\in(0,1).
\end{equation} 

When $\coefB\ll1$, we can rewrite \cref{ma_mass} using \cref{ma_coupling} as follows
\begin{equation}
  \label{eq:ma_mass_b}
  - \frac{\mathrm{d}J_{\conc}}{\mathrm{d}\sx} + \frac{1}{\coefE}\partDeriv{\phi}{t} = 0.
\end{equation}
Using~\cref{vin_pe_small:bc} and setting $J_{in}=1$, we can integrate \cref{eq:ma_mass_b} over the filter depth and time to obtain:
\begin{equation}
  \dhc(t) = \pack\coefE \int_0^t \efficiency(s) \dint s.
\end{equation}
Hence, the dirt-holding capacity $\dhc$ becomes a cumulative measure of the number efficiency $\efficiency$ in filtration regimes when  $\coefB \ll 1$ and the boundary conditions \eqref{vin_pe_small:bc} are used. For this reason, in what follows we mainly discuss the dirt-holding capacity~$\dhc$ and not the number efficiency~$\efficiency$ in all filtration regimes except in the diffusion-only regime. 


\subsection{Advection regime \label{sec:adv}}
If diffusion is negligible in comparison to advection, we have that $\coefC\ll1$ in addition to $\coefB\ll1$ and the time derivative in \cref{ma_mass} being negligible. Then, the Robin boundary condition \eqref{vin_pe_small:bc1} reduces to a Dirichlet boundary condition. We solve \cref{ma_mass,vin_pe_small:bc1} and find the concentration distribution as:
\begin{equation}
  \conc(\sx, t) = \frac{J_{in}}{\coefD\svelv(\sx,t)}\exp\lr{-\int_0^{\sx} \frac{\area(\phi(z,t))}{\coefD\svelv(z,t)} \dint z}.
  \label{red_conc}
\end{equation}
Thus, in the advection-only regime the model simplifies greatly. This is particularly useful for an extended parameter study, where the same problem must be solved many times for different sets of parameters.


\subsection{Diffusion regime \label{sec:diff}}
When diffusion is the dominant transport mechanism of contaminants inside the filter, we neglect the advection ($\coefD\ll1$) and keep all other terms in \cref{ma_mass}. We note that when the diffusion is important the timescale of the contaminant trapping can be comparable with the diffusive processes and so $\coefB$ may not necessarily be small. 

To facilitate the diffusion transport of the contaminants across filter media, a large difference in concentrations at the opposite sides of the media has to be maintained, in contrast to when advection is present. To model this set-up mathematically, we specify Dirichlet boundary conditions for the concentration at both sides of the media:
\begin{equation}
   \label{diff_bc}
   \conc(0, t) = \conc_{in} = 1, \quad \conc(1,t) = 0, \qquad t\in[0,T].
\end{equation}
We note that in this regime, the contaminant transport \cref{ma_mass} decouples from the fluid flow \cref{flow1a,flow2}. Therefore, we do not solve the fluid flow problem in this regime


\subsection{\texorpdfstring{$\svelv_{in}=const$}{} regime \label{sec:vin}}
To specify the constant flow rate, we use the following boundary conditions for the macroscopic equations \eqref{flow1a} and \eqref{flow2}:
\begin{equation}
  \label{vin_bc}
  \svelv(0,t) = \svelv_{in}, \qquad 
  \pres(1,t)  = 0, \qquad
  t\in[0,T].
\end{equation}
The parameter~$\coefA$ in \cref{flow2} represents the ratio between the timescale of the fibre growth due to the contaminant deposition and the timescale of the fluid advection through the filter media, which is usually very small. Using  $\coefA\ll1$ and \cref{vin_bc}, the solution of the one-dimensional flow model~\cref{flow1a,flow2} is given by:
\begin{subequations}
  \label[equation]{vin_sol}
  \begin{alignat}{2}
  \svelv(\sx,t) &= \svelv_{in}, &\sx\in(0,1)\text{ and }t\in[0,T];\\
  \pres(\sx,t) &= \int_{\sx}^1\frac{1}{\sPerm(z,t)}\dint z,\qquad  &\sx\in(0,1)\text{ and }t\in[0,T].
  \end{alignat}
\end{subequations}
We note that the permeability~$\sPerm$ is used only to compute the pressure distribution and it does not affect the velocity, which is constant in space and time. Since the mass-transport problem is coupled with the flow problem only via the fluid velocity (see \cref{ma_mass,ma_coupling}), $\sPerm$ does not impact the mass transport, and as a result the efficiency~$\efficiency$, in this operational regime.


\subsection{\texorpdfstring{$\Delta\pres=const$}{} regime} \label{sec:dp}
To model the filtration regime with a constant pressure drop, we use the following boundary conditions for the flow equations \eqref{flow1a} and \eqref{flow2}:
\begin{equation}
  \pres(0,t) = \Delta\pres, \qquad
  \pres(1,t) = 0, \qquad
  t\in[0,T].
\end{equation}
In this case the solution of the one-dimensional flow problem is:
\begin{equation}
  \svelv(x,t) = \bar{\sPerm}(t)\lr{\Delta\pres
  + \int_0^1 \frac{1}{\sPerm(\phi(y,t))} \lr{\coefA\int_0^{y}\partDeriv{\phi(z,t)}{t}\dint z}  \dint y} - \coefA\int_0^{\sx}\partDeriv{\phi(y,t)}{t}\dint y,
\end{equation}
where
$$
\bar{\sPerm}(t) = \lr{\int_0^1 \frac{1}{\sPerm(\phi(y,t))} \dint y}^{-1}.
$$
Moreover, using again that $\coefA\ll1$, we find 
\begin{equation}
  \label{dp:v}
  \svelv(x,t) = \bar{\sPerm}(t)\Delta\pres, \qquad\sx\in(0,1),\: t\in[0,1].
\end{equation}
Therefore, in since regime we see that the permeability~$\sPerm$ has direct impact on the fluid velocity and thus the efficiency.


\section{Multiscale simulations \label{sec:multi_sim}}


In this section we present simulations of our multiscale model using the five microstructure types from \cref{sec:micro} and the filtration regimes discussed in \cref{sec:macro_method}. In particular, we consider three different scenarios for the transport of the contaminant, diffusive, advective, or both. For the latter two, we distinguish between filtration processes performed under the constant flow rate $\svelv_{in}=const$ or the constant pressure drop $\Delta\pres=const$ (see a summary of each regime in  \cref{tab:regimes}). 

To maintain generality across the examples considered, we use the dimensionless model and, therefore, all parameters and results presented are also dimensionless. The input parameters are presented in \cref{tab:test1}. Below we discuss the simulation results for each filtration regime in detail. A summary of the results in terms of the four sensitivity characteristics $\dhcsens,\pressens,\velsens$, and $\timesens$ is shown in \cref{tab:sensitivity}.

\begin{table*}
	\centering
		\begin{tabular}{lccccc}
            Filtration regime & $\coefA$ & $\coefB$ & $\coefC$ & $\coefD$ & $\coefE$ \\
		    \hline
		    Advection--diffusion & $0$ & $0$ & $1$ & $1$ & $1$ \\
		    Advection & $0$ & $0$ & $0$ & $1$ & $1$ \\
		    Diffusion & $-$ & $0.1$ & $1$ & $0$ & $1$ \\
		\end{tabular}
	\caption{Dimensionless parameters for all filtration regimes.}
	\label{tab:regimes}
\end{table*}

\begin{table*}
	\centering
		\begin{tabular}{lll}
            Parameter & Definition & Value \\
		    \hline
			$\phi_0$ & Initial porosity & $0.93$ \\
            $\delta l/r$ & Micro-lengthscale w.r.t. fibre radius & $3.5$ \\
            $J_{in}$ & Inflow contaminant flux & $1$ \\
            $\conc_{in}$ & Inflow concentration  & $1$ \\            
            $\svelv_{in}$ & Inflow velocity & $1$ \\
            $\Delta \pres_0$ & Pressure drop & $50$ \\
            $\phi_{min}$ & Minimum porosity & $0.5$ \\ 
		\end{tabular}
	\caption{Input parameters for all numerical experiments.}
	\label{tab:test1}
\end{table*}

\begin{table*}
	\centering
		\begin{tabular}{llcccc}
            \multicolumn{2}{l}{Filtration regime} & $\dhcsens$ & $\pressens$ & $\velsens$ & $\timesens$ \\
		    \hline
		    \multirow{2}{*}{$\svelv_{in}=const$} & Advection--diffusion & $0.09$ & $0.58$ & $-$ & $0.2$ \\
		    & Advection & $0.1$ & $0.5$ & $-$ & $0.26$ \\
 		    \hline
		    \multirow{2}{*}{$\Delta\pres=const$} & Advection--diffusion & $0.39$ & $-$ & $0.36$ & $0.33$ \\ 
		    & Advection & $0.48$ & $-$ & $0.39$ & $0.43$ \\ 
 		    \hline
		    Diffusion & & $0.17$ & $-$ & $-$ & $0.26$ \\
		\end{tabular}
	\caption{Characteristics of the microstructure sensitivity for all filtration regimes.}
	\label{tab:sensitivity}
\end{table*}


\subsection{Advection--diffusion and \texorpdfstring{$\svelv_{in}=const$}{} regime \label{sec:diffadv_vin}}
The fluid flow is described by \cref{vin_sol}. The results of the numerical simulations for the five different microstructures are shown in \cref{fig:diffadv_vin}. We observe some variations in the number efficiency~$\efficiency$ and the dirt-holding capacity~$\dhc$ in Figures~\ref{fig:diffadv_vin}A and B, respectively. The random microstructure with no isolation distance and monodisperse fibres has the lowest number efficiency of all microstructures considered. On the other hand, the random microstructure with the polydisperse fibre radii shows the best number efficiency until $t\approx 0.5$, after which the regular microstructures exhibit the largest efficiency. Since the random microstructure with no isolation distance does not store as much contaminant as the other microstructures, it is not surprising that its lifetime is the longest.  However, overall the lifetime of the media is not very sensitive to the microstructure: the  sensitivity characteristic is $\timesens=0.2$ (see \cref{tab:sensitivity}). The variations in the dirt-holding capacity~$\dhc$ for the different microstructures (\cref{fig:diffadv_vin}B) are also small ($\dhcsens = 0.09$). In contrast, the pressure drop~$\Delta\pres$ (\cref{fig:diffadv_vin}C) shows significant variations depending on the microstructure ($\pressens=0.58$). For example, the square grid microstructure reaches a pressure of approximately $160$ at the end of the lifetime of the filter media, while the random microstructure with zero isolation distance reaches only around~$60$. 


\subsection{Advection and \texorpdfstring{$\svelv_{in}=const$}{} regime}
\begin{figure}
	\centering
    \includegraphics[width=0.5\linewidth, page=3]{Fig09}
    \vskip1mm
    \begin{minipage}[t]{.48\textwidth}
	\centering
    \includegraphics[width=1\linewidth, page=2]{Fig09}
    \vskip1mm
    \includegraphics[width=1\linewidth, page=4]{Fig09}
    \vskip1mm
    \includegraphics[width=1\linewidth, page=1]{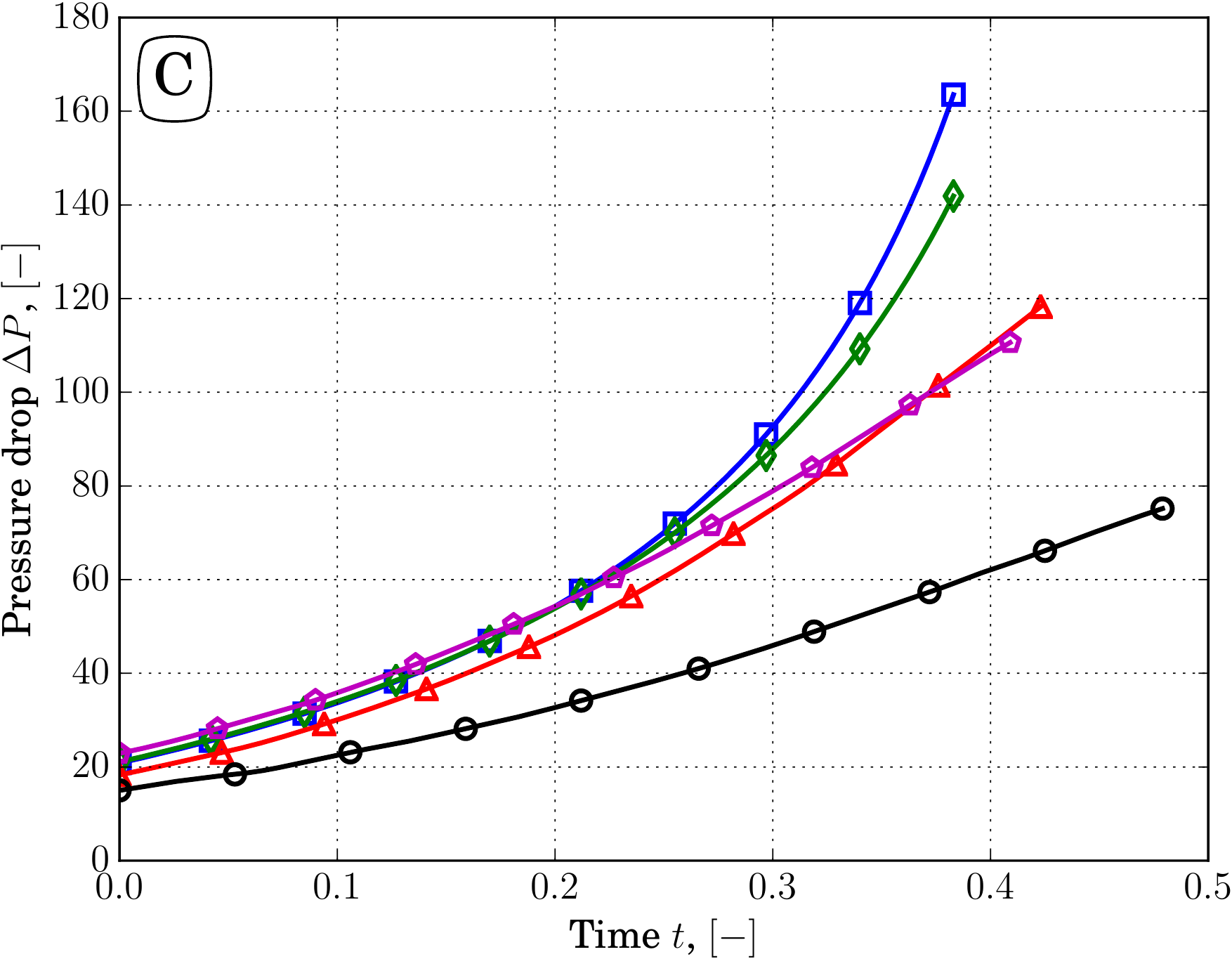}
	\caption{Macroscale simulation results for \textit{advection--diffusion} and $\svelv_{in}=const$ regime. We show the number efficiency~$\efficiency$~(A), the dirt-holding capacity~$\dhc$~(B) and the pressure drop~$\Delta P$~(C) as functions of time $t$.}
    \label{fig:diffadv_vin}
    \end{minipage}
    \hfill
    \begin{minipage}[t]{.48\textwidth}
	\centering
    \includegraphics[width=1\linewidth, page=2]{Fig10}
    \vskip1mm
    \includegraphics[width=1\linewidth, page=4]{Fig10}
    \vskip1mm
    \includegraphics[width=1\linewidth, page=1]{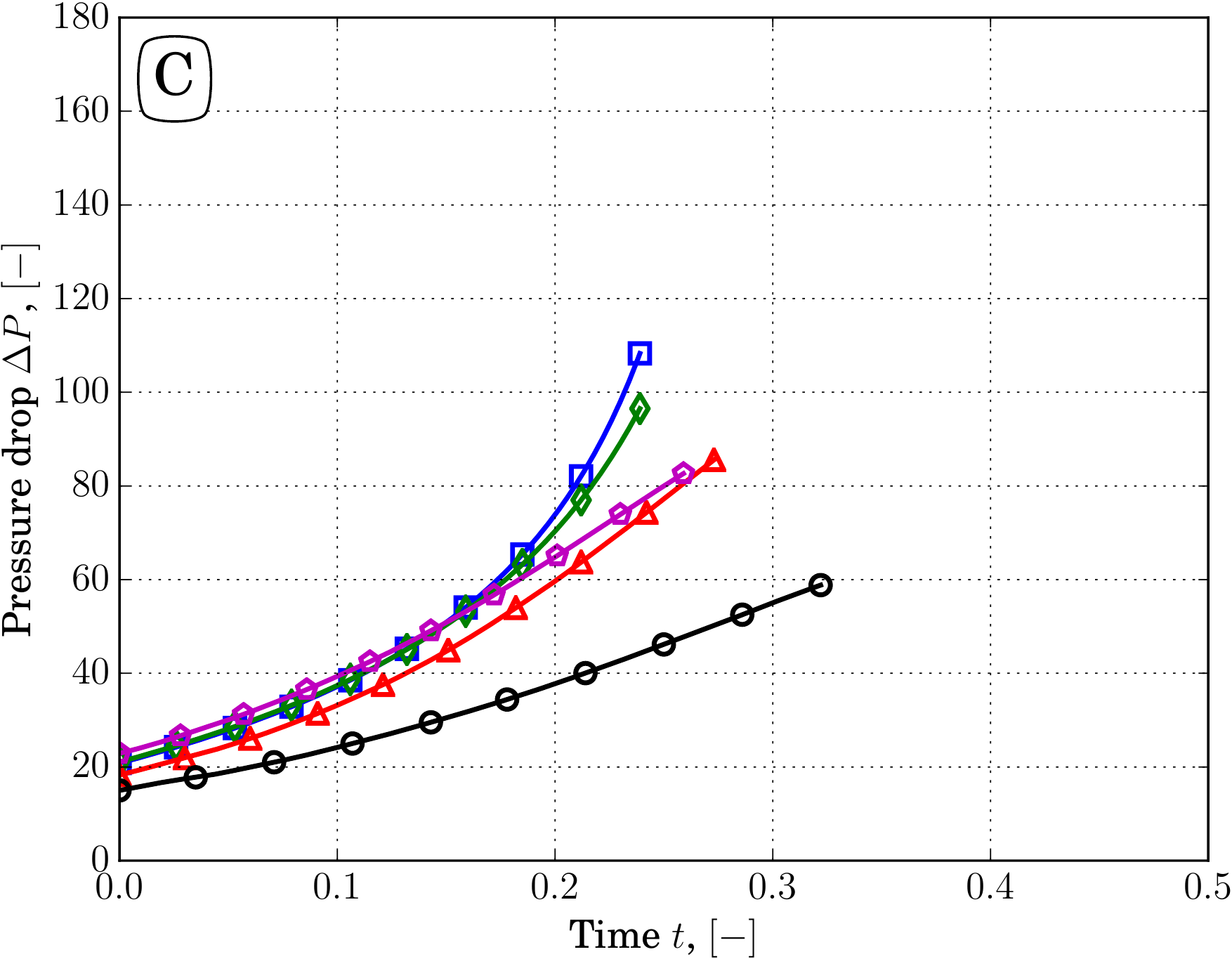}
	\caption{Macroscale simulation results for \textit{advection} and $\svelv_{in}=const$ regime. We show the number efficiency~$\efficiency$~(A), the dirt-holding capacity~$\dhc$~(B) and the pressure drop~$\Delta P$~(C) as functions of time $t$.}
    \label{fig:adv_vin}
	\end{minipage}
\end{figure}

\Cref{fig:adv_vin} shows the simulation results for the advection-only flow regime with  constant flow rate for the different microstructures. We observe that the dirt-holding capacity~$\dhc$ is not significantly influenced by the microstructure type ($\dhcsens=0.1$, \cref{fig:adv_vin}B), while the pressure drop~$\Delta\pres$ depends strongly on the microstructure model ($\pressens=0.5$, \cref{fig:adv_vin}C). The sensitivity of the lifetime of the filter media is $\timesens=0.26$. 

Comparing this regime with the advection--diffusion regime in \cref{sec:diffadv_vin},  we observe that the efficiency and dirt-holding capacity values are higher in the absence of the diffusive transport mechanism. This means that  diffusion reduces the filtration efficacy for the same adsorption coefficient (see discussion in \cref{sec:homog} and further investigations in \cref{sec:transport}). 
The pressure drop in the advection--diffusion regime is more sensitive to the different microstructure types than in the advection regime, while the dirt-holding capacity and the lifetime are slightly less sensitive to the microstructure. Overall, in these two filtration regimes with the constant flow rate, the permeability~$\sPerm$ significantly impacts the pressure drop, but does not influence the efficiency performance for the reasons discussed in \cref{sec:vin}. In contrast, variations across microstructures in the effective diffusivity~$\sDeff$ and the effective surface area~$\area$ have a small effect on the efficiency and dirt-holding capacity. 
For both filtration regimes, the random microstructure with the polydisperse fibre radii shows the best efficiency and dirt-holding capacity while having an average lifetime. Although this microstructure initially has the highest pressure drop, during the lifetime of the filter media  the pressure drop does not increase as rapidly as for the regular microstructures and so when the filter reaches its lifetime the pressure drop is as low as that for the random microstructure with isolation distance~$2r$. 

\subsection{Advection--diffusion and \texorpdfstring{$\Delta\pres=const$}{} regime}
In this filtration regime, the fluid flow is determined by \cref{dp:v}. \Cref{fig:diffadv_dp} shows the simulation results for the different microstructure models. The number efficiency~$\efficiency$ and the dirt-holding capacity $\dhc$ show strong dependence on the microstructure of the filter media in Figures~\ref{fig:diffadv_dp}A and \ref{fig:diffadv_dp}B, respectively. The maximum difference in the dirt-holding capacity arises between the polydisperse random microstructure and the monodisperse random microstructure with no isolation distance, which have the best and worst dirt-holding capacity respectively. The sensitivity of the dirt-holding capacity is characterized by $\dhcsens=0.39$. Again, the filter with the random microstructure with no isolation distance has a longer lifetime than all other microstructures because it does not store as much contaminant. Overall, the lifetime of the filter media is quite sensitive to the microstructure ($\timesens=0.33$). Since the pressure drop is held constant during the filtration, the resulting pressure distribution for all microstructures is the same and is not shown here. Instead, we show the inflow velocity $\svelv_{in}$ as a function of time in \cref{fig:diffadv_dp}C, which demonstrates the throughput of the fluid. The random microstructure with polydisperse fibre radii has the smallest throughput until around $t=0.3$ at which point it switches with the regular square and hexagonal microstructures. The random microstructure with no isolation distance demonstrates the largest throughput for all time. In general, the throughput is sensitive to the microstructure ($\velsens=0.36$).


\subsection{Advection and \texorpdfstring{$\Delta\pres=const$}{} regime}
\begin{figure}
	\centering
    \includegraphics[width=0.5\linewidth, page=2]{Fig11}
    \vskip1mm
    \begin{minipage}[t]{.48\textwidth}
	\centering
    \includegraphics[width=1\linewidth, page=1]{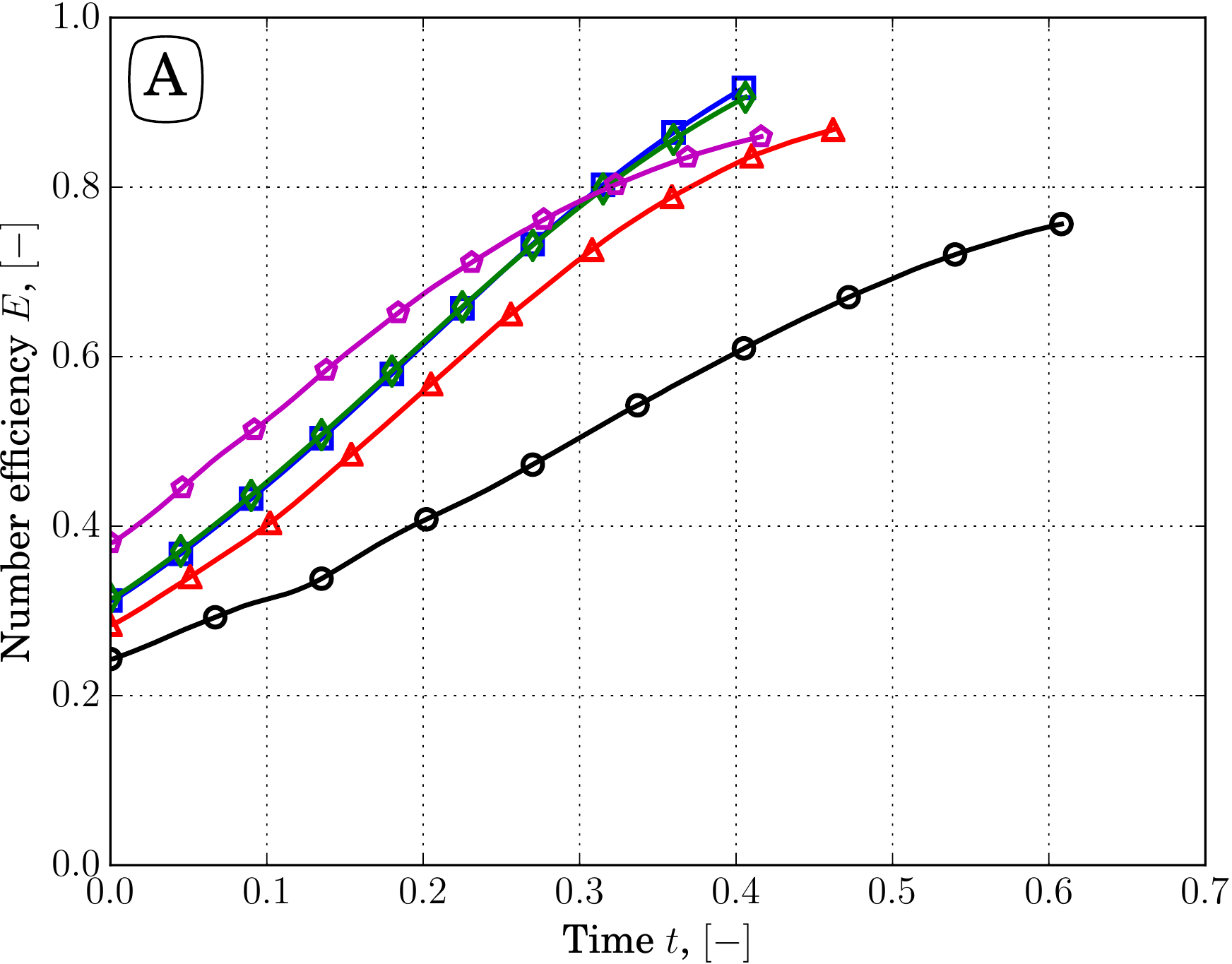}
    \vskip1mm
    \includegraphics[width=1\linewidth, page=3]{Fig11}
    \vskip1mm
    \includegraphics[width=1\linewidth, page=4]{Fig11}
	\caption{Macroscale simulation results for \textit{advection--diffusion} and $\svelv_{in}=const$ regime. We show the number efficiency~$\efficiency$~(A), the dirt-holding capacity~$\dhc$~(B) and the pressure drop~$\Delta P$~(C) as functions of time $t$.}
    \label{fig:diffadv_dp}
    \end{minipage}
    \hfill
    \begin{minipage}[t]{.48\textwidth}
	\centering
    \includegraphics[width=1\linewidth, page=1]{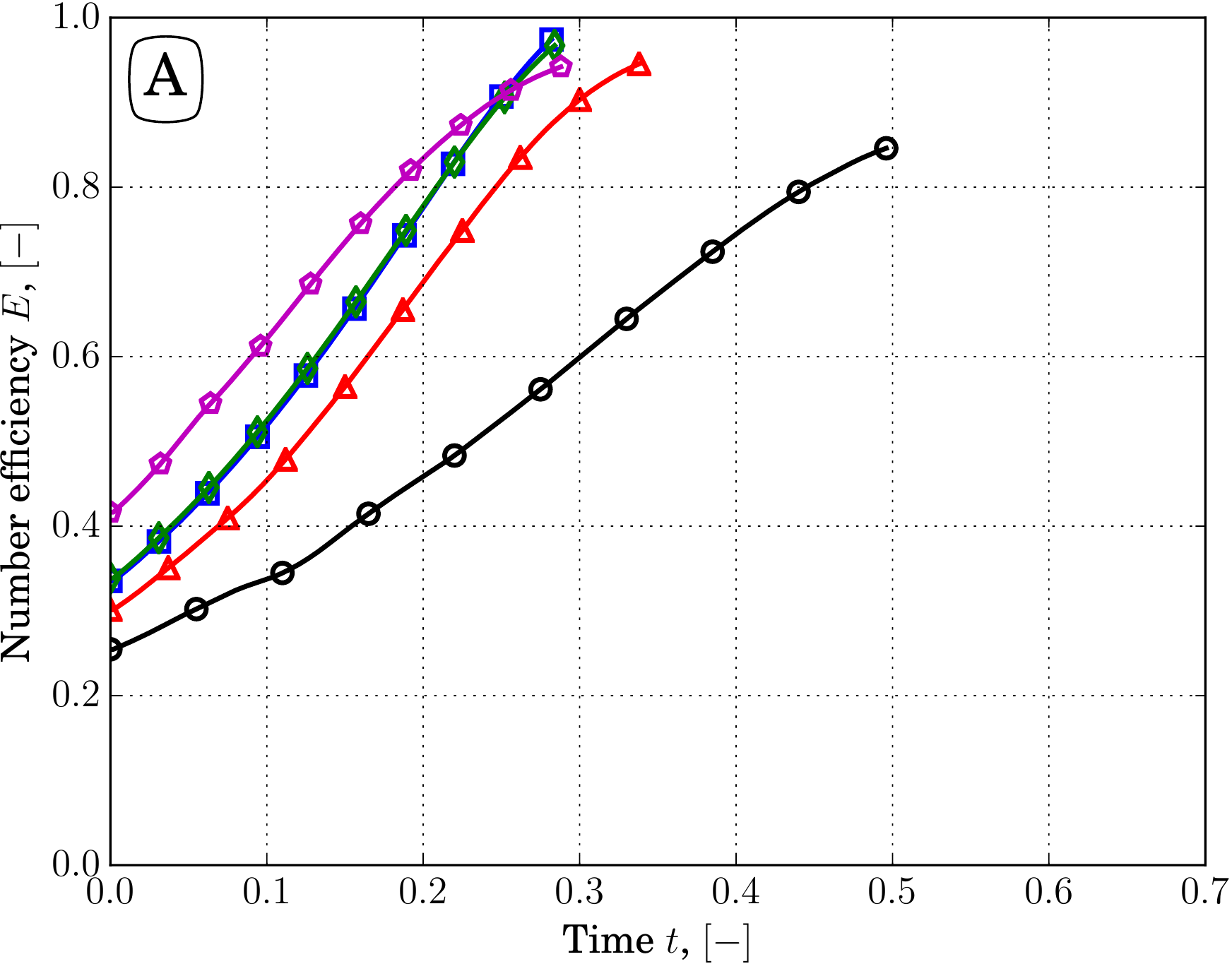}
    \vskip1mm
    \includegraphics[width=1\linewidth, page=3]{Fig12}
    \vskip1mm
    \includegraphics[width=1\linewidth, page=4]{Fig12}
	\caption{Macroscale simulation results for \textit{advection} and $\svelv_{in}=const$ regime. We show the number efficiency~$\efficiency$~(A), the dirt-holding capacity~$\dhc$~(B) and the pressure drop~$\Delta P$~(C) as functions of time $t$.}
    \label{fig:adv_dp}
	\end{minipage}
\end{figure}

This regime is described by \cref{dp:v,red_conc,ma_coupling}. Figures~\ref{fig:adv_dp}A and \ref{fig:adv_dp}B show that the number efficiency~$\efficiency$ and the dirt-holding capacity~$\dhc$ are highly influenced by the microstructure. The sensitivity characteristic for the dirt-holding capacity is $\dhcsens=0.48$, the largest among  the considered filtration regimes. The lifetime of the filter media is between $0.28$ for the random microstructures with polydisperse fibre radii and $0.5$ for the random one with no isolation distance ($\timesens=0.43$). The throughput of the contaminated fluid has also a large sensitivity, $\velsens=0.39$.

The filtration regimes when the pressure drop is held constant (this and previous subsection) experience a large increase in the efficiency during the lifetime of the filter. Moreover, these regimes are influenced significantly by the microstructure of the filter media (see \cref{tab:sensitivity}). This implies that the effective parameters in \cref{ma_mass} have a significant impact on the overall filtration. Comparing the two filtration regimes with the constant pressure drop, we observe that the advection regime is more sensitive to the microstructure than the advection--diffusion regime, which is similar to what we observed for the regimes with constant flow rate. 

In both regimes with constant pressure drop, the random microstructure with the polydisperse fibre radii has the best efficiency and dirt-holding capacity values. However, its lifetime and throughput are significantly lower than those of the random microstructure with no isolation distance. Therefore, if the priority of an application is the efficiency, then the best filter media is one with a random microstructure with the polydisperse fibre radii. If the priority is the lifetime or throughput, then the best one is with the random microstructure with zero isolation distance. 

\subsection{Diffusion regime}

\begin{figure}
	\centering
    \includegraphics[width=0.5\linewidth, page=2]{Fig13}
    \vskip1mm
    \includegraphics[width=0.5\linewidth, page=1]{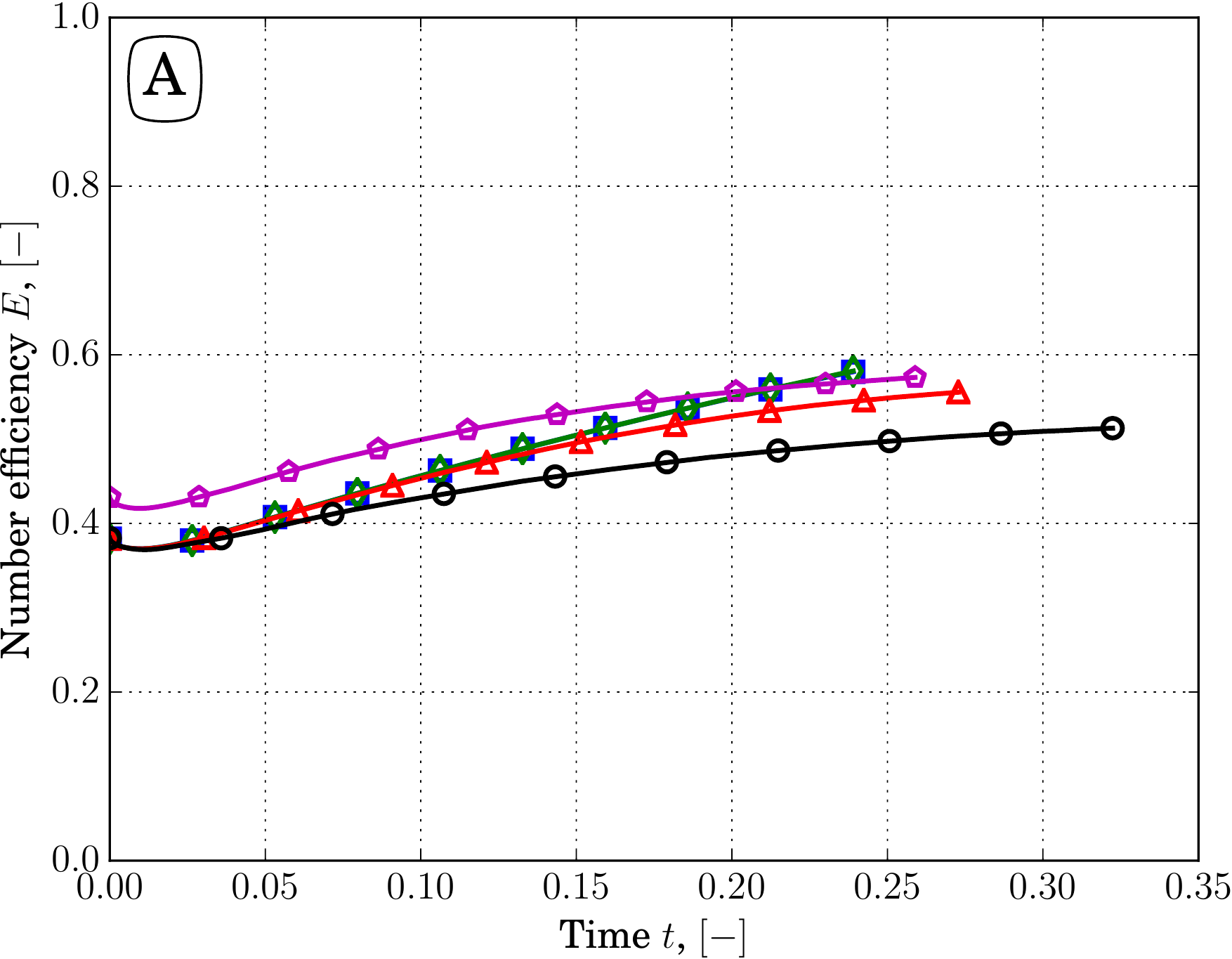}
    \vskip1mm
    \includegraphics[width=0.5\linewidth, page=3]{Fig13}
	\caption{Macroscale simulation results for \textit{diffusion regime}. We show the number efficiency~$\efficiency$~(A) and the dirt-holding capacity~$\dhc$~(B) as functions of time $t$.}
    \label{fig:diff}
\end{figure}

In this regime, we solve \cref{ma_mass,ma_coupling} with boundary conditions \cref{diff_bc} and parameters specified in \cref{tab:regimes}. Figures~\ref{fig:diff}A and \ref{fig:diff}B show the number efficiency~$\efficiency$ and the dirt-holding capacity~$\dhc$, respectively. The sensitivity characteristics for these two metrics are $\dhcsens=0.17$ and $\timesens=0.26$ (\cref{tab:sensitivity}). This regime is more influenced by the microstructure type than filtration regimes with the constant flow rate, but less sensitive to the microstructure than regimes with the constant pressure drop. Similarly to $\Delta\pres=const$ regimes, the best choice of the microstructure depends on the priorities of the filtration: the efficiency or the lifetime.

\section{Transport mechanisms of filtration \label{sec:transport}}
Our simulation results show that including diffusion as a transport mechanism while maintaining a constant advection, the initial efficiency decreases, see for example, \cref{fig:diffadv_vin,fig:adv_vin}. In this section we consider a simplified version of our homogenised model to understand the relative contributions of diffusion and advection. We note we assume the adsorption rate is fixed, although in reality this may change with changes in the contaminant diffusivity.

We consider the initial filtration behaviour before any changes in the porosity affect the microstructure and the filter efficiency. Then, we are concerned with the steady-state version of~\cref{ma_mass}, that is, $\conc(\sx,t=0) \equiv \conc(\sx)$. We assume that the initial microstructure has no macroscopic variations throughout the medium depth, that is, $\phi$, $\sDeff$, $\area$ are constant. The fluid velocity $\svelv$ is also constant because we do not consider changes in the microstructure. We focus on the case when the contaminant influx is prescribed, see \cref{vin_pe_small:bc}. Then, the filtration model \cref{ma_mass} in 1D reads:
\begin{subequations}
\label[equation]{eq:initmass}
   \begin{alignat}{2}
    -a\frac{\mathrm{d}^2\conc}{\mathrm{d}\sx^2} 
    + b\frac{\mathrm{d}\conc}{\mathrm{d}\sx} + \conc &= 0, 
    \qquad &\sx\in(0,1);\\
    -a \frac{\mathrm{d}\conc}{\mathrm{d}\sx} + b \conc &= c,
    &\sx = 0; \\
    \frac{\mathrm{d}\conc}{\mathrm{d}\sx} &= 0,
    &\sx = 1.
   \end{alignat}
\end{subequations}
where the parameters $a$, $b$ and $c$ are defined as follows:
\begin{alignat}{3}
a=\frac{\coefC\phi\sDeff}{\area},\qquad & 
b=\frac{\coefD\svelv}{\area},\qquad & 
c=\frac{J_{in}}{\area}.
\end{alignat}
The solution of the \cref{eq:initmass} is given by
\begin{equation} \label{con_exact}
  \conc(x) = K_1 \exp\lr{\lambda_1\sx} + K_2 \exp\lr{\lambda_2\sx},
\end{equation}
where:
\begin{align}
\lambda_{1,2} &= \frac{b \pm \sqrt{b^2 + 4a}}{2a},\\
  K_i &= c\lr{b - a\lambda_i 
  - \frac{\lambda_i\exp(\lambda_i)}{\lambda_j\exp(\lambda_j)}
  \lr{b - a\lambda_j}}^{-1},
  \quad i,j=1,2, \quad i\neq j.
\end{align}
The initial number efficiency is, from \cref{eq:eta},
\begin{equation}
  \label{eq:eff}
  \efficiency(t=0) \equiv \efficiency_0 = 1 - \frac{b}{c} \conc(1).
\end{equation}
We find that the time $t^*$ that a single contaminant particle takes to transit the filter is: 
\begin{equation}
  \label{eq:time}
  t^* = \int_0^1 \frac{\conc(\sx)}{J_{\conc}(\sx)} \dint\sx.
\end{equation}

\begin{figure}
  \begin{minipage}[t]{.48\textwidth}
  \centering
  \includegraphics[width=1\linewidth, page=1]{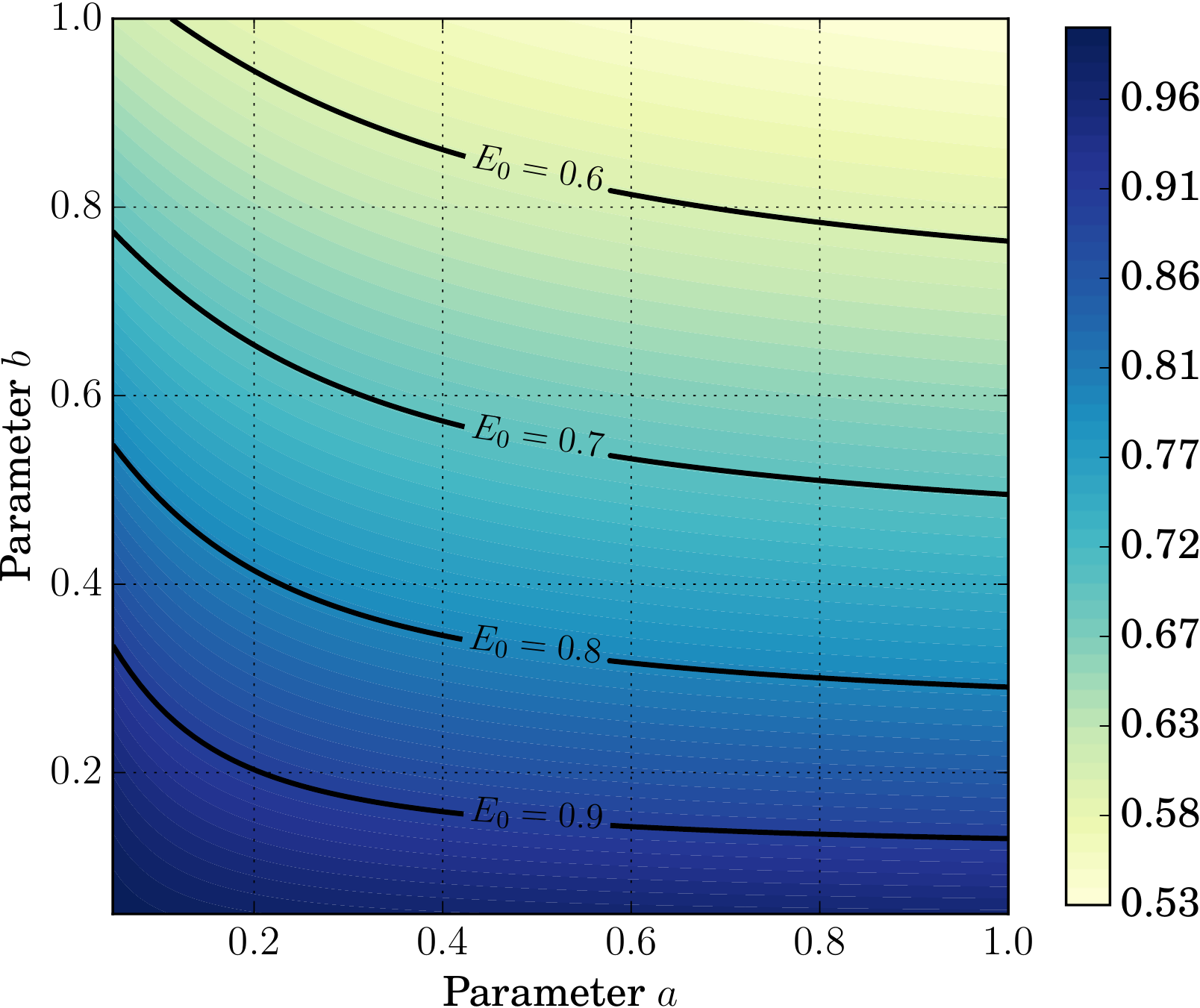}
  \caption{The initial efficiency $\efficiency_0$ is shown as function of the parameters $a$ and $b$.}
  \label{fig:initeff}
  \end{minipage}
  \hfill
  \begin{minipage}[t]{.48\textwidth}
  \centering
  \includegraphics[width=1\linewidth, page=2]{Fig14-15}
  \caption{The time $t^*$ taken by a contaminant particle to transit the filter is shown as function of the parameters $a$ and $b$.}
  \label{fig:transtime}
  \end{minipage}
\end{figure}

Since we are interested in the contributions of the diffusion and advection transport mechanisms, which are represented by the parameters $a$ and $b$ respectively, let us assume that $c=1$.  
\Cref{fig:initeff} shows the initial efficiency $\efficiency_0$ as function of $a$ and $b$. This corroborates our simulation results that as diffusion gets smaller, the efficiency increases. Similarly, the efficiency increases when we reduce the advection. Overall, the behaviour of the number efficiency is monotone with respect to the advection and diffusion terms. 

\cref{fig:transtime} shows the transit time $t^*$ as a function of the parameters $a$ and $b$. A larger diffusion~$a$ means that, for the same concentration gradient, the transport of the contaminant will be faster. Therefore, as $a$ increases, the transit time~$t^*$ reduces along with the chance for this particle to adhere to the fibre surface, which yields lower efficiency. The same holds for the increased advection~$b$ and we observe monotone decrease of the transit time as $a$ and $b$ increase. 

This offers a route towards improving the filtration efficiency, and quantifying the improvements gained, by adjusting parameters of the filtration set-up. However, we note that these conclusions are made based on the assumption of a constant adsorption coefficient, while in real filtration processes this  adsorption might change with space and time as discussed earlier.


\section{Conclusions \label{sec:conc}}


In this study, we obtained a multiscale model using the method of multiple scales to simulate contaminant filtration in fibrous filter media with unidirectional fibres. Our main contribution was to study how sensitive filter performance is to the microstructure under different filtration regimes. First, we extended the homogenization model by \citet{DBG15,DBG16} to account for random microstructures and fibres with different sizes in the same unit cell. Second, we proposed an agglomeration algorithm to model the process whereby fibres that are located close to one another agglomerate as contaminant deposits on their surface. 

In filtration processes, it is important to be able to account for nonuniform filter porosities: these appear in porosity-graded filters, but also in standard uniform filters that become porosity-graded as contaminant is deposited in a nonuniform way in its depth. In our model, we accounted for nonuniform porosity by allowing the microscopic cell geometry to vary with the macroscopic variable. The advantage of this approach is that we were able to parameterize the nonuniform microstructure via the porosity $\phi$ and compute the effective parameters in the homogenized model (the permeability~$\sPerm$, the effective diffusivity~$\sDeff$, and the effective surface area~$\area$) as a function of $\phi$ as a pre-processing step. In other words, there is a one-way coupling between the microscopic model (the cell problems) and the macroscopic homogenized model describing the evolution of the contaminant concentration across the filter depth. The disadvantage is that it requires the microstructure to be locally periodic, so large variations in porosity are not allowed. This is in contrast with the more general work by \cite{Ray15}, which requires the cell problems to be solved at every macroscopic location. 

Thanks to the one-way coupling, our model provides an efficient simulation tool that can handle regular microstructures (for example, square and hexagonal), but also for random microstructures (random arrangements of fibres in a unit cell that are extended periodically). We investigated five types of microstructures: square, hexagonal, two random with different isolation distances between fibres and random with polydisperse fibre radii. These microstructures provide us with different degrees of randomness. The hexagonal microstructure is the limiting case of a random microstructure with large isolation distance (as the hexagonal lattice maximizes the separation distance between fibres), while the purely random one is with zero isolation distance. The microscale simulations showed that the permeability~$\sPerm$ and the effective surface area~$\area$ are very sensitive to the microstructure of the fibrous media, but the effective diffusivity~$\sDeff$ is less so.

We considered five different filtration regimes and investigate how sensitive the final filtration results to the microstructure type are. We summarize our findings below. 
\begin{itemize}
\item  \textit{Advection--diffusion and advection regimes} with $\svelv_{in}=const$: The best choice for the microstructure is random with polydisperse fibre radii. This set-up provides the best number efficiency and dirt-holding capacity among the microstructures considered, and its pressure does not rise very much during the whole lifetime of the filter media. However, while an optimum microstructure exists, the performance in this regime is only weakly affected by the microstructure choice, and all other microstructures perform almost as well (see \cref{tab:sensitivity}).
\item \textit{Advection--diffusion and advection regimes} with $\Delta\pres = const$: The random microstructure with polydisperse fibre radii also provides the best filtration efficiency and the dirt-holding capacity in these cases. Unless we are interested in the lifetime and throughput of the filter media, then the best microstructure is the random one with no isolation distance. Filtration is a lot more sensitive to the microstructure in these regimes.
\item \textit{Diffusion regime}: Again, the random microstructure with polydisperse fibre radii exhibits the best efficiency results, while the random microstructure with no isolation distance has the longest lifetime. This regime has a moderate sensitivity to the microstructure in comparison with the other filtration regimes.
\end{itemize}

To understand the interplay between the transport mechanisms in the filtration, we investigated how the diffusion and advection terms affect the efficiency results. We found that the initial number efficiency decreases monotonically as we increase the strength of their dimensionless groups. This behaviour is explained by the fact that larger  diffusion or advection lead to a faster transport of contaminant  across the filter media and consequently less time for them to adhere to the fibre surface.
 
Our multiscale model can be used to make predictions about filter performance and find optimize its porosity or microstructure depending on the requirements of the application. This could range from air-conditioning systems, pharmaceutical and biotechnology industries.   Predicting experimental data with a mathematical tool always introduces additional challenges: defining realistic parameters that cannot be measured (for example, the adsorption coefficient) and accounting for additional features of real-life processes (for example, polydisperse contaminants and parameter uncertainties). Moreover, we will also extend this model to account for additional effects that can influence the filtration performance, such as electrostatic effects in air filtration. This will require deriving a new mathematical model accounting for more physical complexity while having the similar objective of developing a simple and efficient tool.

In the wider context, the mathematical framework that we have laid out applies to a range of other problems in which obstacle growth and coalescence are important. For example, in the application of tissue engineering and cell growth, the model may be used to understand the effect that irregularities in the scaffold have on the rate of tissue growth. The problem may also be studied in reverse, whereby material is being removed from rather than deposited on obstacles. Such a scenario may occur in various geological applications with dissolution of porous media or in cases where the goal is to remove a particular substance, such as in the decontamination of chemical from a porous medium~\citep{dalwadi2017mathematical}. The model could also be generalized to account for transfer of material from obstacle to obstacle, mimicking scenarios such as Ostwald ripening~\citep{Voorhees1985}.

\section*{Acknowledgements}
This publication was made possible by an EPSRC Impact Acceleration Account, grant EP/K503769/1, and research funding from Dyson Ltd. The authors would like to thank the Dyson staff, in particular Gareth Morris, Ben Hovell, Tom Grimble and Stefan Koch, for inspiring this research project and for fruitful discussions on the problem understanding and research directions. IMG is grateful to the Royal Society for a University Research Fellowship. MB is grateful to St John's College, Oxford, for funding through a Junior Research Fellowship.


\appendix


\section{Convergence test for closely located fibres \label{app_1}}
\begin{figure}
  \centering
  \includegraphics[width=0.5\linewidth, page=1]{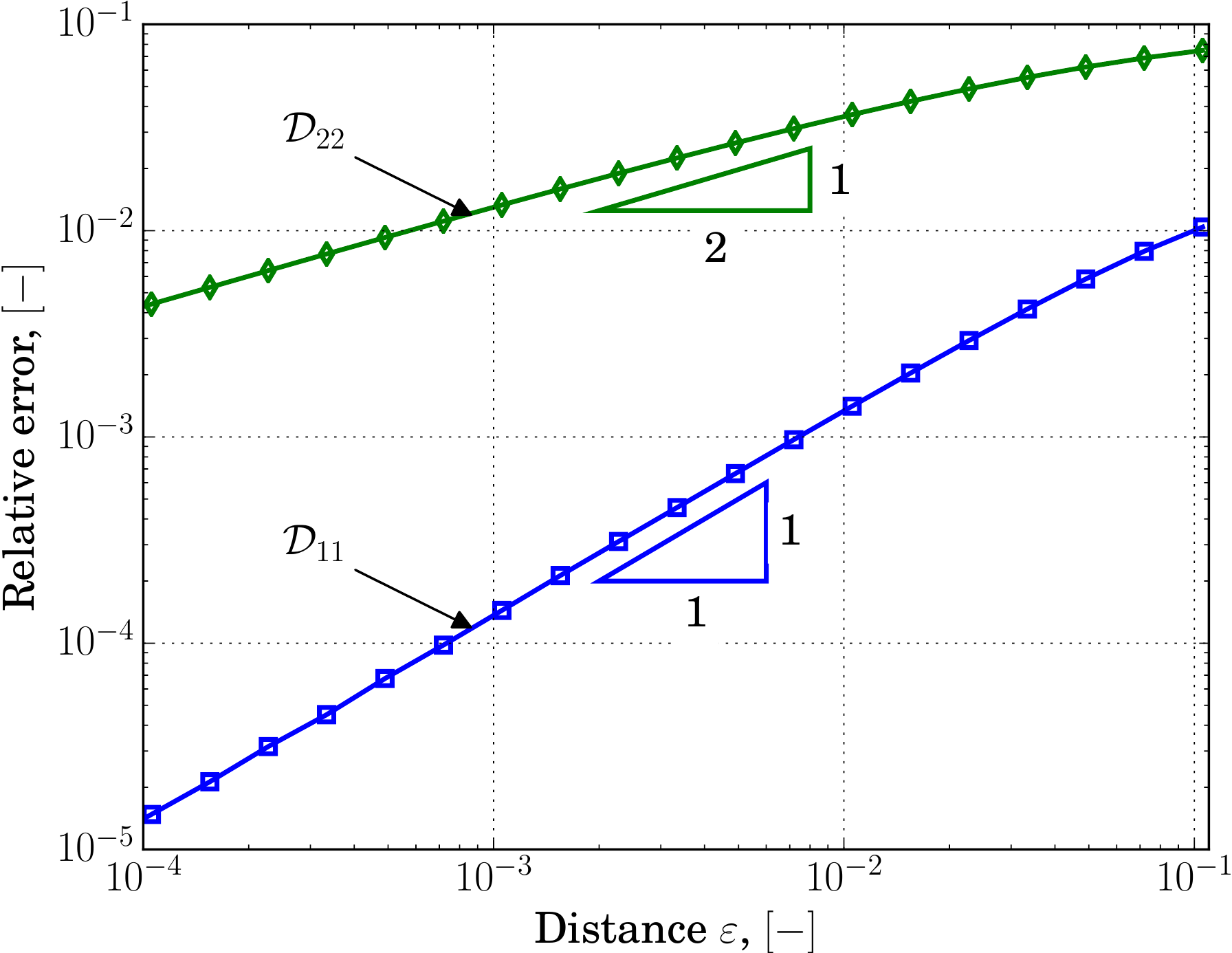}
  \caption{Relative error between two close fibres and the limiting case of two touching fibres as a function of the distance between two close fibres.}
  \label{fig:conv_plot}
\end{figure}

As the distance between two fibres tends to zero, the diagonal elements of the Jacobian matrix of $\mathbf{\Gamma}$, which is a solution to the cell problem \cref{mass_cell}, tend to infinity in a domain whose measure tends to zero. To make sure that this numerical effect does not cause problems while computing the effective diffusivity $\deff$ for the random microstructures, we perform a convergence test. We consider a unit cell with two fibres placed at the centre of a unit cell with fibre centres located on the same horizontal line. We denote the minimal distance between their surfaces in horizontal direction as $\varepsilon$. To obtain a reference solution, we use a unit cell with the limiting case, that is, when the two fibres are just in contact and form a connected `infinity-shaped 'volume. Then, we compute a relative error of the effective diffusivity as a function of $\varepsilon$ (see \cref{fig:conv_plot}). 

The estimation of the relative error requires a refinement around the fibres, which poses a restriction on the smallest size of the critical distance that we can consider. We are able to estimate the error for the critical distances $\varepsilon$ up to $10^{-4}$ (see \cref{fig:conv_plot}), which show a monotonic convergence of the effective parameters for the two closely located fibres to the limiting case with two joined fibres. The relative error for the diagonal elements of the effective diffusivity $\deff$ is less than $10^{-2}$ for the critical distance $\varepsilon = 10^{-4}$. This distance is usually less than average size of a contaminant particles and it is likely to be blocked by a single particle, which yields to the formation of an agglomerate in a real filtration process. Hence, forming the agglomerates for the distances around $10^{-4}$ in the modelling process not only produces small errors, but is also justified from a physical standpoint.

\providecommand{\noopsort}[1]{}\providecommand{\singleletter}[1]{#1}%

\end{document}